%% file: SailFFish_Preprint.tex
\documentclass{article}

\usepackage{PRIMEarxiv}

\usepackage[utf8]{inputenc} 
\usepackage[T1]{fontenc}    
\usepackage{hyperref}       
\usepackage{url}            
\usepackage{booktabs}       
\usepackage{amsmath}        
\usepackage{amsfonts}       
\usepackage{nicefrac}       
\usepackage{microtype}      
\usepackage{lipsum}
\usepackage{fancyhdr}       
\usepackage{graphicx}       
\graphicspath{{media/}}     

\usepackage{pgfplots}
\pgfplotsset{compat=newest} 
\newlength\figureheight
\newlength\figurewidth 

\usepackage{multirow}

\hypersetup{
    colorlinks=true,
    linkcolor=blue,
    filecolor=magenta,      
    urlcolor=cyan,
    pdftitle={Overleaf Example},
    pdfpagemode=FullScreen,
    }

\title{SailFFish: A Lightweight, Parallelised Fast Poisson Solver Library
}

\author{
  Joseph Saverin \\
  Hermann Föttinger Institute \\
  Technische Universität Berlin \\
  Berlin, Germany\\
  j.saverin@tu-berlin.de \\
}

\begin{document}
\maketitle

\begin{abstract}
A solver for the Poisson equation for 1D, 2D and 3D regular grids is presented. The solver applies the convolution theorem in order to efficiently solve the Poisson equation in spectral space over a rectangular computational domain. Conversion to and from the spectral space is achieved through the use of discrete Fourier transforms, allowing for the application of highly optimised $\mathcal{O}(N\log N)$ algorithms. The data structure is configured to be modular such that the underlying interface for operations to, from and within the spectral space may be interchanged. For computationally demanding tasks, the library is optimised by making use of parallel processing architectures. A range of boundary conditions can be applied to the domain including periodic, Dirichlet, Neumann and fully unbounded. In the case of Neumann and Dirichlet boundary conditions, arbitrary inhomogeneous boundary conditions may be specified. The desired solution may be found either on regular (cell-boundary) or staggered (cell-centre) grid configurations. For problems with periodic, Dirichlet or Neumann boundary conditions either a pseudo-spectral or a second-order finite difference operator may be applied. For unbounded boundary conditions a range of Green's functions are available. In addition to this, a range of differential operators may be applied in the spectral space in order to treat different forms of the Poisson equation or to extract highly accurate gradients of the input fields.
The underlying framework of the solver is first detailed, followed by a range of validations for each of the available boundary condition types. Finally, the performance of the library is investigated. The code is free and publicly available under a GNU v3.0 license at \href{https://github.com/ZeppSav/SailFFish}{github.com/ZeppSav/SailFFish}.
\end{abstract}

\keywords{Poisson Solver \and Fast Fourier Transform \and Spectral Solver, Numerical Integration}

\section{Introduction}
\label{sec:intro}
For any given vector field $\vec{v}$ which demonstrates $C^2$ smoothness-- implying that the second derivative (or higher) is continuous-- the field may be expressed using the Helmholtz decomposition:
\begin{equation}
\label{eq:helm}
\vec{v} = \nabla\times\vec{\psi} - \nabla\phi 
\text{  ,}\hspace{5mm}\text{ where } \hspace{5mm}
\nabla\cdot\vec{\psi} = 0
\text{  .}
\end{equation}
Here $\vec{\psi}$ is a vector potential which represents the solenoidal (divergence-free) part of the field and $\phi$ is a scalar potential which represents the irrotational part of the field. The divergence free constraint on the vector potential is necessary to ensure a unique decomposition. By taking the curl of \eqref{eq:helm}, one retrieves the Poisson equation: a linear, second order partial differential equation (PDE): 
\begin{equation}
\label{eq:pois}
\nabla^2 \vec{\psi} = \vec{f} 
\text{  ,   }
\end{equation}
where $\vec{f} = -\nabla\times\vec{v}$ is the curl of the vector field. By taking the divergence of \eqref{eq:helm}, one again retrieves a Poisson equation relating the Laplacian of the scalar potential and the divergence of the vector field $\nabla\cdot\vec{v}$. This direct connection to continuous vector fields illustrates precisely why the Poisson equation is so ubiquitous in physical modelling. The conservation of physical field quantities leads to vector and scalar potentials which are amenable to solution with the Poisson equation or simplifications thereof. Applications include fluid dynamics, electrostatics, electromagnetism, gravitation, diffusion and heat transfer, amongst many others. A range of methods exist for the solution of the Poisson equation depending on the desired accuracy of the solution and boundary conditions (BCs) of the domain. These may be broken down broadly into three categories: multigrid methods, fast multipole methods and spectral methods \cite{Gholami_2016}. For the work here spectral methods have been applied, however the other solution methods shall briefly be described for comparison.

\subsection{Multigrid Methods}
Multigrid (MG) methods can be applied for the solution of a range of both nonlinear and linear PDEs on arbitrary grid configurations over a range of domain topologies \cite{Brandt_1982, Brandt_1990, Henson_2003, Venner_2000}. The main concept of the MG method is to solve the desired PDE on a range of grid resolutions $\mathcal{H}_{i=0,1,\cdots n}$. This shall be described first for a two grid levels $\mathcal{H}_0$, $\mathcal{H}_1$. An initial solution is found on a disjoint portion of the finer grid using e.g. finite differences. A residual error is calculated by comparing source and preliminary solution. This residual is anterpolated (adjoint interpolation) from the base grid $\mathcal{H}_0$ to $\mathcal{H}_1$. The solution is then found on the coarser grid and interpolated back down to the finer grid to improve the initial guess. This gives rise to a iterative process of progressively smoothing the solution until a convergence criteria is satisfied. This process is generally extended to higher grid levels to further improve efficiency. Numerous schemes are available for optimally traversing between grid levels e.g. $V$ and $W$-cycles depending on the problem type \cite{Adams_1989}. It can be demonstrated that for a grid with $N$ grid points, the computational expense scales as $\mathcal{O}(N)$ \cite{Brandt_2001}. An additional advantage of such methods is their amenability to irregular domains or domains with variable resolution.
\subsection{Fast Multipole Methods}
The Fast Multipole Method (FMM) was initially developed by Greengard \& Rokhlin \cite{Greengard_1988, Greengard_1997}. This method makes use of the Green's function solution to the Poisson equation:
\begin{equation}
\label{eq:green}
\psi(r) = \mathcal{G}*f = \int_{\mathcal{R}^n} f(r')\,\mathcal{G}(r-r')\,d^nr 
\text{  ,}
\end{equation}
where ($*$) indicates the convolution operator. The Green's function $\mathcal{G}$ represents the fundamental solution to the unbounded Poisson equation: $\nabla^2\mathcal{G}(r) = \delta (r)$ where $\delta (r)$ is the Dirac delta function \cite{Kreyszig_Book}. FMM exploits the nature of $\mathcal{G}$ to create a low-rank approximation to the far-field influence. $\mathcal{G}(r)$ may be expressed as a multipole expansion which, in the far-field requires a smaller expansion for a desired accuracy $\epsilon$. The use of a hierarchical grid allows coarser levels of approximation with larger interaction distances. 
The difficulty in formulating the multipole expansion of the desired kernel motivated development of the kernel-independent FMM (KIFMM) \cite{Ying_2004} or similar methods such as the multi-level multi-interaction cluster (MLMIC) \cite{VanGarrel_2017, Saverin-2018-AIAA}. An advantage of the FMM is that it naturally handles unbounded BCs, although periodic or homogeneous Neumann or Dirichlet BCs may also be treated by using the method of images \cite{Cocle_2008, Saverin_PhD}. A significant advantage of the FMM is the ability to handle sparse or even disjoint source distributions. Typically the FMM scales either as $\mathcal{O}(N\log N)$ or $\mathcal{O}(N)$ depending on problem and setup \cite{Greengard_1988}.

\subsection{Spectral Methods}
\label{sec:spec}
The third class of methods exploit the linearity of the Poisson equation by representing the solution as a sum of basis functions. It shall be assumed without loss of generality that in the case of interest here, these are trigonometric functions. The forward and inverse Fourier transforms of a function are introduced:
\begin{equation}
\label{eq:ft}
\text{Forward: } \hspace{5mm} 
\hat{f}(k) = \int_{-\infty}^{\infty}f(x)e^{-2\pi ikx}\,dx
\text{,} \hspace{10mm} 
\text{Inverse: } \hspace{5mm} 
f(x) = \frac{1}{2\pi}\int_{-\infty}^{\infty}\hat{f}(k)e^{2\pi ikx}\,dk
\text{ , }
\end{equation}
This transforms the representation of the function from real ($x$) space to spectral ($k$) space. The Poisson equation may be written in the spectral space as:
\begin{equation}
\label{eq:pois_spec}
-k^2\hat{\psi}(k) = \hat{f}  
\text{  ,}\hspace{5mm}\text{ where } \hspace{5mm}
k^2 = 
\begin{cases}
k_x^2   & \text{in 1D} \\
k_x^2 + k_y^2   & \text{in 2D} \\
k_x^2 + k_y^2 + k_z^2  & \text{in 3D}
\end{cases} \text{  ,}
\end{equation}
and $k_i$ are the angular wave numbers in the spectral space. Applying the convolution theorem, convolution in the real space is represented in the spectral space by point-wise multiplication. Herein lies the enormous advantage of this approach. The appropriate Green's function \eqref{eq:green} need only be calculated once, converted to the spectral representation and stored. Thereafter the entire convolution of any given source field is equivalent to a multiplication in the spectral space. 
In practice, the continuous transforms of \eqref{eq:ft} are treated with a discrete Fourier transform (DFT) which represents the Fourier transform in a discrete, truncated spectral space. Application of the DFT allows one to exploit the fast Fourier transform (FFT) to achieve a computational expense $\mathcal{O}(N\log N)$ \cite{Cooley_1965}. Although this appears to not be a significant improvement over MG or FMM approaches, the myriad applications of the FFT in signal processing implies that highly optimised libraries exist for the calculation of FFTs. Unlike the MG or FMM approaches whereby a desired accuracy of the solution $\epsilon$ is achieved, the use of the spectral approach allows, under suitable conditions, spectral accuracy to be achieved. This implies the solution is practically exact, with the error saturating at machine precision. A range of BCs may be represented including Neumann, Dirichlet, periodic and unbounded. These shall be detailed in the proceeding section. The method has the disadvantage that the entire grid must be generated and stored for a given resolution which leads to large memory overheads. This makes the method less suitable for sparse source distributions. In addition, extensions of the method to non-rectangular domains or irregular grids can be challenging \cite{Poplau_2003}. \newline 

For the work carried out here, the spectral method has been implemented within SailFFish, an open-source library written in \texttt{C++}. The library has been written in such a way that emphasis is placed on simplicity, modularity and adaptability. The code is available for free online under a GNU v3.0 license at \href{https://github.com/ZeppSav/SailFFish}{github.com/ZeppSav/SailFFish}. The rest of the paper is devoted to an overview and validation of the SailFFish solver. In Section~\ref{sec:method} the numerical method implemented in the solver is detailed. In Section~\ref{sec:framework} the framework of the solver is described including details on the application of object-oriented programming, data type modularity and input/output interfaces. In Section~\ref{sec:validation} a range of validation cases have been investigated to demonstrate the accuracy of the solver. In Section~\ref{sec:perf} library performance is detailed along with scaling behaviour. Finally in Section~\ref{sec:conc} conclusions are drawn and a scope for future features and applications is described. 

\section{Method of Solution}
\label{sec:method}
The spectral method has been applied in SailFFish for the solution of the Poisson equation. The methodology and solution steps are described in this section along with the options for different solver types. For illustrative purposes a simple 1D periodic case will first be described. As detailed in Section \ref{sec:spec}, it shall be assumed that the function is expressed as a trigonometric series:
\begin{equation}
\label{eq:trig_series_sol}
f(x) = \sum_{k=0}^{N} \hat{f}_ke^{2\pi i\frac{kx}{L_x}} 
= \sum_{k=0}^{N} \
\hat{f}_k\left(\cos 2\pi k\frac{x}{L} + i\sin 2\pi k\frac{x}{L}\right)
\text{ ,}
\end{equation}
where $L$ is the length of the domain. The periodicity of the solution over the domain can be deduced from this representation. The trigonometric functions represent solution eigenvectors.
By substituting this representation into \eqref{eq:pois}, one obtains the following representation of the solution on the grid:
\begin{equation}
\label{eq:spect_sol}
-\left(\frac{2\pi k}{L}\right)^2\hat{\psi}_k = \hat{f}_k
\text{ .}
\end{equation}
The factor multiplying the $\hat{\psi}_k$ represents the spectral eigenvalues corresponding to the solution eigenvectors \cite{Fuka_2015}. Regarding the solution accuracy, the (backwards) transformed values $\psi_n$ on the grid will correspond exactly to the solution of the Poisson equation, to machine precision. This is referred to as demonstrating  \emph{spectral} accuracy. It is observed that the case $k=0$ represents a division by zero. The Fourier mode $k=0$ represents the average value of the function $\psi$ and taking this to be zero leads to a zero mean and avoids numerical instabilities. \newline 
An alternative approach to the solution of the Poisson equation is to directly apply the representation \eqref{eq:trig_series_sol} with finite differences (FD) \cite{Wiegmann_1999}. Unlike the accuracy achieved with the spectral eigenvalues described above, this approach converges with the accuracy of the FD scheme applied. \newline 
A third approach is to calculate the Green's function \eqref{eq:green} in real space, and then transform this to retrieve the spectral space representation. This approach has the advantage that mollified Green's kernels may be applied which allow for smoothing of the input field \cite{Caprace_2021,Hejlesen_PhD,Hejlesen_2019}. This approach has been applied to unbounded solver cases in SailFFish {\textendash} see Section \ref{sec:unbounded}.\newline 

\subsection{Transformation between Real and Spectral Space Using DFTs}
\label{sec:transforms}
As opposed to the continuous forward and inverse Fourier transforms as defined by \eqref{eq:ft}, in practice the grid values are specified on a discrete, equispaced grid. The conversions to and from the discrete spectral space are achieved with the forward and inverse DFT, defined as:
\begin{equation}
\label{eq:dft}
\text{Forward: } \hspace{5mm} 
\hat{f}_d(k) = \sum_{n=0}^{N-1}f_d(x)e^{-2\pi ik\frac{n}{N}}
\text{,} \hspace{10mm} 
\text{Inverse: } \hspace{5mm} 
f(x) = \frac{1}{N} \sum_{n=0}^{N-1}\hat{f}_d(k)e^{2\pi ik\frac{n}{N}}
\text{ , }
\end{equation}
Depending on the BCs being applied, along with the type of the input data (purely real or complex), different numerical forms of the DFT may be chosen in order to improve performance. As an example, whenever the input data is an array of $n$ purely real values $f = [f_1, f_2, \dots ,f_n]$, the output of the DFT can be shown to demonstrate Hermitian redundancy. A real-to-complex (R2C) DFT may be applied which expresses the output vector as a complex array with $\frac{n}{2}+1$ elements. This simplification achieves approximately a factor of two improvement in speed and memory overhead. The inverse transform is achieved in a similar manner. \newline
Depending on the dimension of the problem being solved, 1D, 2D or 3D DFTs are executed. The 1D DFT is applied to a single, contiguous array of data. Higher dimensional DFTs are applied by carrying out the DFT sequentially in 1D-pencils. A 2D DFT carried out on a set of data points $f(x_i,y_j)$ is therefore calculated as follows:
\begin{enumerate}
    \item \textbf{X forward transform}: Forward DFT taken in $x$ direction to transform $f(x_i,y_j)$ into $\hat{f}(k_x,y_j)$,
    \item \textbf{Y forward transform}: Forward DFT taken in $y$ direction to transform $\hat{f}(k_x,y_j)$ into $\hat{f}(k_x,k_y)$.
\end{enumerate}
The inverse procedure follows precisely the same template, however applying the inverse DFT and proceeding in the reserve order. In the case of the aforementioned 2D example, the inverse transform is executed as follows:
\begin{enumerate}
     \item \textbf{Y inverse transform}: Backward DFT taken in $y$ direction to transform $\hat{f}(k_x,k_y)$ into $\hat{f}(k_x,y_j)$,
    \item \textbf{X inverse transform}: Backward DFT taken in $x$ direction to transform $\hat{f}(k_x,y_j)$ into $f(x_i,y_j)$.
\end{enumerate}
There are two approaches for handling these procedures in 2D. In the first approach, the partially transformed data after step (1) is re-aligned (or stored in an intermediate array) such that it is contiguous in memory for application of the second DFT. This has the advantage of allowing more flexibility in choosing BCs in each spatial direction \cite{Caprace_2021}. The second approach makes use of devoted 2D DFT algorithms which automatically apply the DFTs to the correctly ordered data based on the specified grid dimensions. The latter approach is applied in SailFFish, as will be detailed in Section~\ref{sec:framework}. The extension to 3D cases is straightforward. 
\begin{figure}
    \begin{minipage}{0.45\textwidth}
    \includegraphics[]{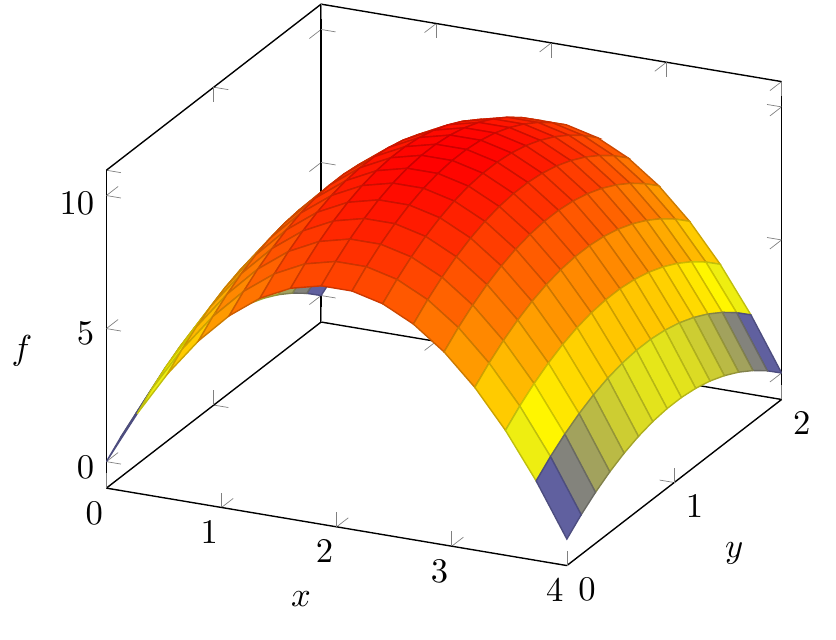}
    \end{minipage}
    \hspace{10mm}
    \begin{minipage}{0.45\textwidth}
    \vspace{4mm}
    \includegraphics[scale=1.08]{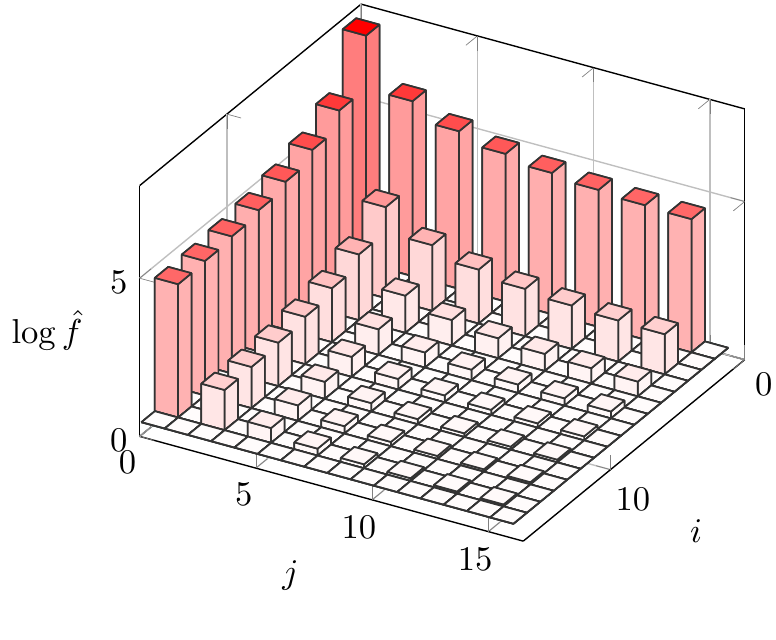}
    \end{minipage}
    \caption{A 2D paraboloid distribution of the source term $f$ (left) along with its representation $\hat{f}$ in the spectral space (right). The logarithm of $\hat{f}$ is shown purely for visual purposes.}
    \label{fig:forwardFT}
\end{figure}

\subsection{Calculating and Retrieving the Solution}
Within SailFFish the spatial convolution is carried out in the spectral space as described in Section \ref{sec:spec}. This is always achieved by  carrying out four steps:
\begin{enumerate}
    \item \textbf{Forward Transform}: The source distribution $f$ is transformed to a spectral representation $\hat{f}$ via a forward DFT;
    \item \textbf{Convolution}: Point-wise multiplication of $\hat{f}$ with the spectral space multiplier (Either the corresponding eigenvalue or the Fourier transform of the Green's function $\hat{\mathcal{G}}$);
    \item \textbf{Spectral Operations}: If necessary, additional operations within the spectral space are carried out;
    \item \textbf{Backward Transform}: The spectral solution  $\hat{\mathcal{G}}\hat{f}$ is transformed to the real space $f$ via an inverse DFT.
\end{enumerate}
The general procedure is therefore relatively straightforward. A visualisation of the outputs of step 1) are shown in Figure \ref{fig:forwardFT}. Despite the simplicity of the above approach, there are a range of solver options. \textbf{Type of DFT}: This is determined by the form of the data (purely real, complex) being transformed and which type of BCs are being enforced (R2C, R2R, DFT). \textbf{Spectral space convolution kernel}: These can either be derived from the spectral or finite difference approaches as described in the previous section, or by applying solutions of the Green's kernel \eqref{eq:green}. \textbf{Spectral space operator}: A range of operators may be applied in the spectral domain to modify the form of the Poisson equation under investigation or to extract spectral gradients. These options are detailed in Section \ref{sec:difops}. \newline
\subsection{Solver Options in SailFFish}
The specification of the solver type and solver setup in SailFFish depends on the desired solution field, BCs and grid properties. The range of options available are described in this section.  
\subsubsection{Solver Types}
\label{sec:solvertypes}
Currently SailFFish has the ability to resolve either 1D/2D/3D scalar fields or 3D vector fields. In the case of a 3D vector field, the three components are resolved by individual component-wise application of the corresponding solver methodology for a 3D scalar solver. Keeping this is mind, four solver classes are available:
\begin{itemize}
    \item \texttt{Poisson\_Periodic\_1D}/\texttt{2D}/\texttt{3D}/\texttt{3DV}: Periodic BCs are applied.
    \item \texttt{Poisson\_Dirichlet\_1D}/\texttt{2D}/\texttt{3D}/\texttt{3DV}:
    Dirichlet (odd function extension) BCs are applied.
    \item \texttt{Poisson\_Neumann\_1D}/\texttt{2D}/\texttt{3D}/\texttt{3DV}:
    Neumann (even function extension) BCs are applied.
    \item \texttt{Poisson\_Unbounded\_1D}/\texttt{2D}/\texttt{3D}/\texttt{3DV}:
    The problem is treated with free-space BCs.
\end{itemize}
In all cases, the \texttt{3DV} option represents the case of a 3D vector field, otherwise scalar fields are resolved. The first three options described above represent \emph{bounded} solver types {\textendash} see Section \ref{sec:bounded}. The final solver type is treated using a different solution method and is separately described in Section~\ref{sec:unbounded}. Currently only uniform BCs are available within SailFFish. For details of an implementation which allows for arbitrary BCs, the reader is referred to the FLUPS library \cite{Caprace_2021}. The reason for this is elaborated upon in Section \ref{sec:framework}. 

\subsubsection{Grid Type}
SailFFish solves the Poisson equation \eqref{eq:pois} on regular grids in 1D, 2D and 3D cases. The resolution of the grid is always specified by the number of grid cells, $N_c$. During solver initialization the upper and lower spatial coordinate for each grid dimension must be specified. Additionally, one has the option to specify the position on the grid where the solution is being found. If the \texttt{REGULAR} grid option is specified, the solution will be calculated at the cell boundary positions. The number of resolved grid position $N_s$ changes depending on the type of solver being applied {\textendash} see Section \ref{sec:bounded}. If the \texttt{STAGGERED} grid option is specified, the solution is found at the centre of the cells. In this case the solution is found at $N_c$ positions. Numerically, the grid positions are given by:
\begin{equation}
\label{eq:gridpos}
X_i = 
\begin{cases}
X_l + i(X_u-X_l)/N_c  & 
\text{for } i=0,1,\dots N_s \hspace{10mm} \text{\texttt{REGULAR} grid option} \\
X_l + (i+0.5)(X_u-X_l)/N_c  & 
\text{for } i=0,1,\dots N_c-1 \hspace{4mm} \text{\texttt{STAGGERED} grid option}
\end{cases}
\text{ , }
\end{equation}
where $X_l$ and $X_u$ are the lower and upper values of the grid domain, respectively. This scheme is applied for each of the resolved spatial directions.  
\subsubsection{Differential Operators}
\label{sec:difops}
In some cases it may be desirable to apply spectral differentiation to the transformed source field $\hat{f}$. Inspecting \eqref{eq:ft}, one observes:
\begin{equation}
\label{eq:ft_sd}
\frac{d}{dx}f(x) = 
\frac{1}{2\pi}\int_{-\infty}^{\infty}\frac{d}{dx}\left(\hat{f}(k)e^{2\pi ikx}\right)\,dk =
\frac{1}{2\pi}\int_{-\infty}^{\infty}ik\,\hat{f}(k)e^{2\pi ikx}\,dk
\text{ , }
\end{equation}
The process of multiplying by $ik$ in the frequency domain hence has the result of extracting the (spectrally accurate) derivatives of the function $f$. 
This may also be applied with a convolution kernel $\hat{\mathcal{G}}(k)$ in order to modify the form of the Poisson equation being solved. A range of options exist in SailFFish to extract the spatial gradients:
\begin{itemize}
\item \texttt{DIV}: Resolves the divergence of the input field
\item \texttt{GRAD}: Resolves the gradients of the input field
\item \texttt{CURL}: Resolves the curl of the input field
\item \texttt{NABLA}: Resolves the Laplacian of the input field
\end{itemize}
The application of spectral differentiation to the point-wise solution to PDEs is commonly referred to as the Pseudo-spectral method \cite{Orszag_1972}. 
\subsection{Bounded Solvers} 
\label{sec:bounded}
If either the nature or the numerical value of the BCs on the domain are specified, the solution on the domain shall be referred to as a \textbf{bounded} solution. Three forms of bounded solvers can be specified: \textbf{Dirichlet} (odd symmetry), \textbf{Neumann} (even symmetry), or \textbf{periodic}. The choice of transform types and their corresponding eigenvalues are described below.
\subsubsection{Transform Types}
The Dirichlet and Neumann type solvers allow the use of a real-to-real (R2R) DFT. Purely real forms of \eqref{eq:trig_series_sol} are utilised whereby only the $\sin$ (discrete sine transform {\textendash} DST) or $\cos$ (discrete cosine transform {\textendash} DCT) expansions are applied for symmetric odd or symmetric even BCs, respectively. Depending on whether a regular or staggered grid option is chosen also dictates the solution eigenvectors and therewith the type of DST or DCT being applied. These are summarised for all boundary condition and grid type variations in Table \ref{tab:Bounded}. Further references are provided here for a detailed description of the corresponding eigenvector functions \cite{Fuka_2015,Caprace_2021,FFTW05,FFTW97}. 
By nature of the fact that it is composed of periodic trigonometric functions, a DFT is itself periodic. For periodic BCs the direct complex-to-complex (C2C) standard DFT is applied. Although this may be simplified for purely real data by applying the R2C transform described in Section \ref{sec:transforms}, this is configured for complex inputs for generality for future application cases. Real data inputs are stored as complex data inputs with imaginary component zero. 
\begin{table}[ht]
 \caption{DFT Type and Eigenvalues for Bounded Solver Types}
  \centering
  \begin{tabular}{|cccccc|}
    \toprule
\multicolumn{1}{|p{2cm}|}{\centering BC \\ Type }
& \multicolumn{1}{|p{2cm}|}{\centering Grid \\ Type }
& \multicolumn{1}{|p{2cm}|}{\centering DFT \\ (Forward) }
& \multicolumn{1}{|p{2cm}|}{\centering DFT \\ (Backward) }
& \multicolumn{1}{|p{2cm}|}{\raggedleft $\mathcal{G}_i$ \\ (PS) }
& \multicolumn{1}{|p{2cm}|}{\centering $\mathcal{G}_i$ \\ (FD2) } \\
    \hline \hline
Dirichlet   & Regular  & DST-I & DST-I     & $-\left(\pi\frac{k+1}{L}\right)^2$  
                                            & $-\frac{4}{H^2}\sin^2 \frac{\pi(k+1)}{2(N+1)}$ \\
Dirichlet   & Staggered & DST-II& DST-III   & $-\left(\pi\frac{k+1}{L}\right)^2$  
                                            & $-\frac{4}{H^2}\sin^2 \frac{\pi (k+1)}{2N}$ \\
Neumann     & Regular   & DCT-I & DCT-I     & $-\left(\pi\frac{k}{L}\right)^2$    
                                            & $-\frac{4}{H^2}\sin^2 \frac{\pi k}{2(N-1)}$ \\
Neumann     & Staggered & DCT-II & DST-III  & $-\left(\pi\frac{k}{L}\right)^2$    
                                            & $-\frac{4}{H^2}\sin^2 \frac{\pi k}{2N}$ \\
\multicolumn{1}{|p{2cm}}{\centering Periodic }
& \multicolumn{1}{p{2cm}}{\centering Both }
& \multicolumn{1}{p{2cm}}{\centering DFT }
& \multicolumn{1}{p{2cm}}{\centering Inverse DFT }
& \multicolumn{1}{p{3.5cm}}{\centering $-\left(2\pi\frac{k}{L}\right)^2$: k $<\frac{N}{2}$ \\ 
                                     $-\left(2\pi\frac{N-k}{L}\right)^2$: k $\geq\frac{N}{2}$ }
& \multicolumn{1}{p{2cm}|}{\centering $-\frac{4}{H^2}\sin^2 \frac{\pi k}{N}$ } \\
    \bottomrule
    \end{tabular}
    \label{tab:Bounded}
\end{table}

\subsubsection{Choice of Eigenvalues}
\label{sec:coe}
\begin{figure}
    \hspace{-10mm}
    \begin{minipage}{0.35\textwidth}
    \includegraphics[scale=1.0]{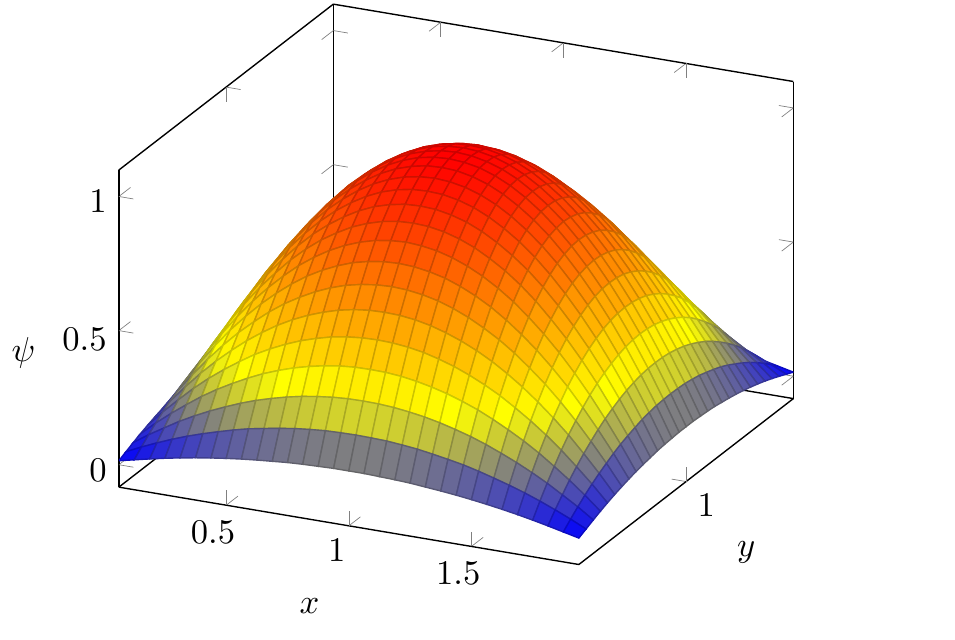}
    \end{minipage}
    \begin{minipage}{0.35\textwidth}
    \hspace{30mm}
    \includegraphics[scale=1.0]{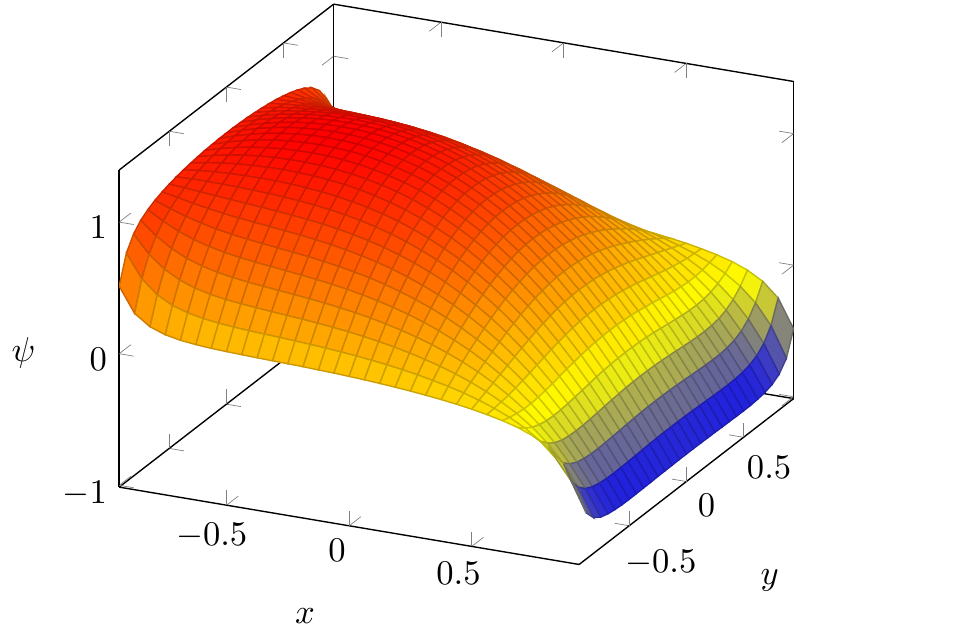}
    \end{minipage}
    \caption{The solution to a 2D Dirichlet problem with homogeneous (left) and inhomogeneous (right) boundary conditions using FD2 type eigenvalues. The source field $f$ is equivalent in both cases.}
    \label{fig:dirichlet}
\end{figure}

For bounded problems, two options for the eigenvalues are available. The first are pseudo-spectral (PS) eigenvalues. As was illustrated above for the periodic case, there are derived directly from the eigenvalue resulting from differentiation in the real space of the trigonometric expansion. These vary due to the corresponding zeros of the eigenvectors for particular BC-grid type combinations. The second option is the use of the second-order FD (FD2) eigenvalues. Use of this option has the advantage that it allows for the use of inhomogeneous BCs for Dirichlet and Neumann solver types. This is a result of the fact that the finite differences are built into the spectral solution. In order to enforce the desired boundary values a simple weighting of the boundary source terms may be applied \cite{Wiegmann_1999}. This is automatically accounted for in SailFFish based on the solver and grid type. An illustration of inhomogeneous BCs is given in Fig.~\ref{fig:dirichlet}.  
The correct eigenvalues to all available BC and grid type combinations are also provided in Table \ref{tab:Bounded}.
In order to ensure the  periodicity of the desired solution for spectral eigenvalues the application of a cyclic Green's function is required. 
This can be observed for the periodic solvers PS eigenvalue, which is mirrored about the point $k=\frac{N}{2}$. 

\subsection{Unbounded Solvers}
\label{sec:unbounded}
\begin{figure}[ht]
    \hspace{-10mm}
    \begin{minipage}{0.35\textwidth}
    \includegraphics[scale=1.0]{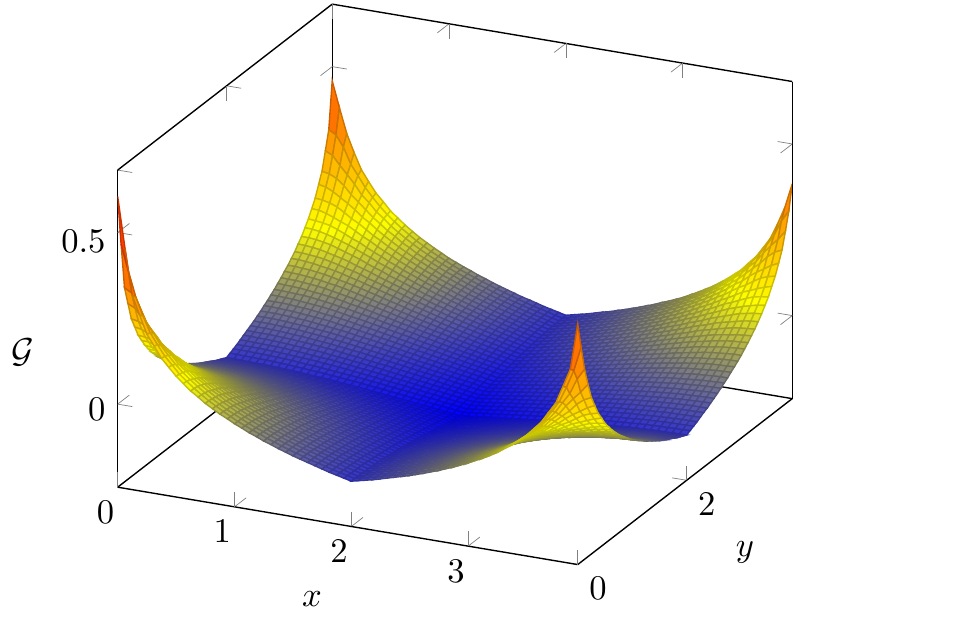}
    \end{minipage}
    \begin{minipage}{0.35\textwidth}
    \hspace{30mm}
    \includegraphics[scale=1.0]{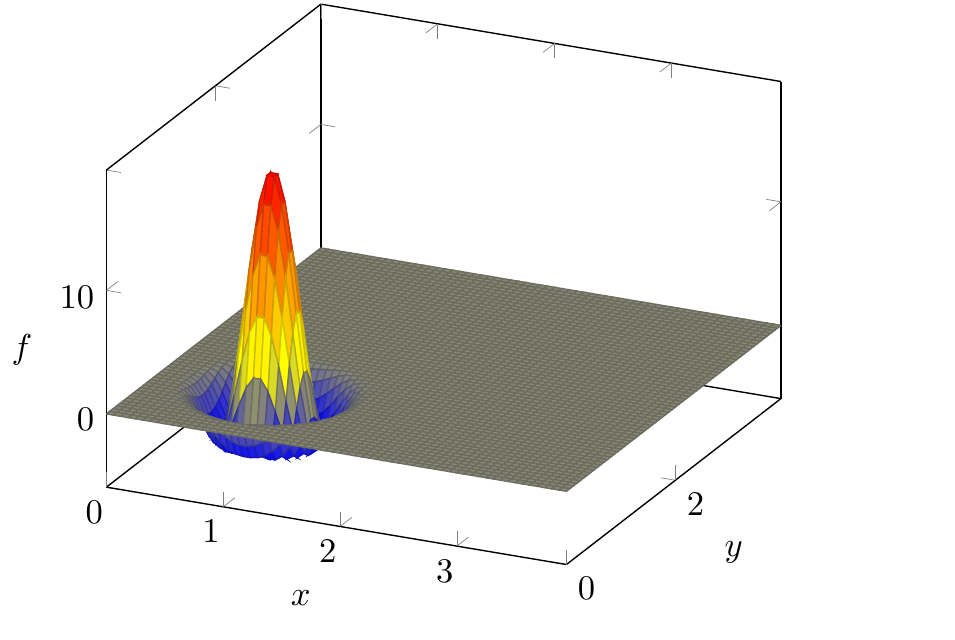}
    \end{minipage}
    \caption{Operations necessary for application of the Hockney-Eastwood method. The doubling and mirroring of the Green's function in real space is carried out (left). The source domain is also doubled and the remaining space is zero-padded (right). For the case here a bump function is being simulated \textendash see Section~\ref{sec:valunb}.}
    \label{fig:unbounded}
\end{figure}

If free-space BCs are specified, the solution on the domain shall be referred to as an \textbf{unbounded} solution. There are numerous approaches to handling this problem including the approach due to James \& Lackner \cite{Lackner_1976,James_1977}, the Vico-Greengard method \cite{Vico_2016}, or the Hockney-Eastwood method  \cite{Hockney_1988,Eastwood_1979}. The latter has been applied in SailFFish. In order to mimic the effect of free boundary conditions, the domain of the Green's function is doubled in the dimensions of interest and mirrored about the point $N=k$, as was done for the periodic spectral eigenvalues. Due to the periodic nature of the DFT, this has the effect that the convolution of any point in the frequency domain extends to the boundary as it would for the unbounded solution. In order to ensure that the convolution in the frequency domain occurs without reflection, the source field must also be doubled, and the additional solver regions are zero-padded to avoid any artificial influence. A visualisation of this is given in Fig.~\ref{fig:unbounded}. After the periodic solution has been found on the doubled domain, the solution within the desired region is extracted and the remaining region is discarded. For an elegant explanation of this concept in 1D and for mixed BCs, the reader is referred to the FLUPS documentation \cite{Caprace_2021}.
\subsubsection{Transform Types}
For unbounded cases, the direct DFT is applied. The source function and Green's function may therefore be complex values. Although not yet implemented in SailFFish, this allows the solution of the Helmholtz equation \cite{Chatelain_2010,Caprace_2021}. In the case that purely real data is investigated, the R2C and C2R transforms are applied to improve solver efficiency. 
\subsubsection{Greens Functions}
Unlike the bounded solvers whereby the Green's functions are available for direct application in the frequency space, more care must be taken for unbounded solutions. The problem of the undefined Green's function at the origin cannot by circumvented. In addition, the original approach of Hockney \& Eastwood only displays $\mathcal{O}(h^2)$ convergence, where $h$ is the grid size \cite{Eastwood_1979}. In order to overcome this, a range of options have been suggested. Hejlesen et al. suggested a range of Gaussian smoothings which progressively approximate the singular solution $\nabla^2\mathcal{G} = \delta (r)$ by applying $P^{th}$ order approximations to the inverse Gaussian regularisation \cite{Hejlesen_PhD}. The same author derived a spectrally compact kernel which achieves spectral accuracy on finite grids \cite{Hejlesen_2019}. Within SailFFish the spectrally compact Hejlesen kernels are available for 1D, 2D and 3D cases. For 2D and 3D solvers the Hejlesen Gaussian approximations of order $P=2, 4, \dots 10$ may also be selected.

\subsection{Semi-unbounded Solvers}
It is possible to manually break 2D or 3D DFT operations down to a set of successive 1D DFT processes. This is the approach taken in a range of other Poisson solver packages \cite{Caprace_2021, HejlesenSolver_Github}. This approach has the advantage that the Laplace operator may be split in the desired direction, yielding a 2D Helmholtz equation which allows for the treatment of mixed boundary conditions \cite{Chatelain_2010}. By applying combinations of DSTs, DCTs and DFTs this allows for the implementation of fully arbitrary boundary conditions \cite{Caprace_2021}.
Within SailFFish the optimised $N$-dimensional DFT algorithms are applied to reduce communication times and allow simple development on a graphical processing unit (GPU) {\textendash} as explained in Section \ref{sec:framework}. This feature is therefore currently not available in SailFFish, however it is planned to implement this in future work.

\section{Solver Framework}
\label{sec:framework}
SailFFish has been prepared with the following objectives in mind:
\begin{enumerate}
    \item Simplicity, readability and abstraction;
    \item Optimise for shared-memory platforms;
    \item Minimise communication time and maximise application of existing libraries and algorithms.
\end{enumerate}
Abstraction has been exploited with the object-oriented structure of \texttt{C++} to avoid code repetition and improve readability. Class objects and inheritance are described in the proceeding section. Preparing the software in this way simplifies tailored application to specific problems for future applications. SailFFish is designed to be run on shared-memory devices because, put simply, not every user has access to cluster computing equipment and the overarching concept should be \textbf{efficiency} rather than power. For all computationally demanding tasks, the multi-platform shared-memory parallel programming framework \texttt{OpenMP} has been utilised to parallelise calculations \cite{OpenMP}. \newline

The final point refers to exploitation of existing optimised programs or APIs in the interest of maintaining readable code. Take for example the execution of 2D and 3D FFTs. If the software is configured to handle mixed BCs, then it is necessary to carry out the transform into and out of spectral space individually in 1D arrays. This requires reorganisation of data such that it is contiguously aligned for the 1D FFT algorithms. This necessity both complicates the code (particular for arbitrary boundary conditions) and increases memory communication overhead as described in \cite{Caprace_2021}. In SailFFish this is avoided by directly exploiting 2D and 3D FFT algorithms which exist within the chosen FFT engine. One advantage of this approach is that data may be passed as very large monolithic blocks. An additional advantage is that operations on the data (e.g. convolution) may be carried out along a single index. \newline

This approach is particularly relevant for the solution of unbounded problems {\textendash} see Section \ref{sec:unbounded}. The Hockney-Eastwood algorithm requires the doubling of the domain in each dimension, leading to an increase in calculation expense and memory overhead by a factor of approximately $2^d$ where $d$ is the dimension of the problem \cite{Hockney_1988}. This can be reduced to a factor of approximately $2^{d-1}$ by carrying out the transforms in each direction sequentially and discarding the additional information in the zero-padded region \cite{Hockney_1988,Hejlesen_PhD}. This necessarily greatly reduces the size of the FFTs. As described above however it also requires reordering the data for the FFT algorithms. This is avoided for now by simply treating the entire doubled domain in order to adhere to the mantra of the simplification. It is the hope of the author that in the near future the necessity to double the domain will be avoided by making direct use of the James-Lackner method to treat the unbounded problem \cite{James_1977,Lackner_1976}. The proceeding sections are devoted to a description of the architecture of the SailFFish library.

\subsection{Architecture of SailFFish}
\begin{figure}[ht]
    \begin{picture}(150,150)
    \put(0,0){\includegraphics[scale=1.2]{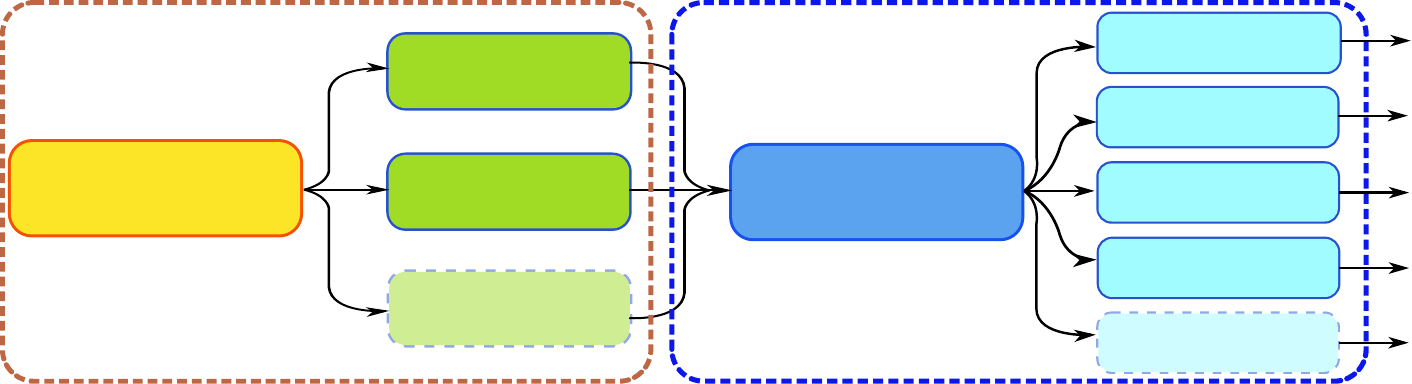}}
    \put(18,100){\underline{DataType Classes}}
    \put(18,65){DataType Parent}
    \put(142,105){FFTW DataType}
    \put(142,65){cuFFT DataType}
    \put(142,25){--User-defined--}
    \put(270,100){\underline{Solver Classes}}
    \put(270,65){Solver Parent}
    \put(400,115){1D\_Scalar}
    \put(400,90){2D\_Scalar}
    \put(400,65){3D\_Scalar}
    \put(400,38){3D\_Vector}
    \put(390,12){--User-defined--}
    \end{picture}
    \caption{Architecture of SailFFish}
    \label{fig:Arch}
\end{figure}
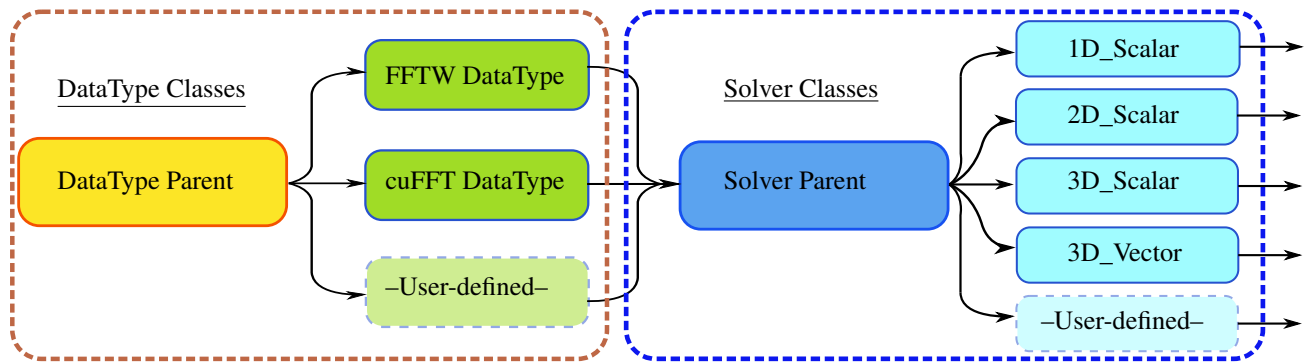

There are two parent classes within SailFFish which contain the majority of the functionality: the \texttt{DataType} class and the \texttt{Solver} class. These are described here. A visualisation of the class data structure is given in Fig.~\ref{fig:Arch}. 

\subsubsection{DataType Class}
The \texttt{DataType} parent class contains a range of virtual methods relevant to accessing (input/output), manipulating (convolution) and carrying out transforms of data within the spectral space. For a given choice of FFT tool, a corresponding class is derived from the \texttt{DataType} parent class and the corresponding virtual methods must be implemented. Currently two \texttt{DataType} bases are available in SailFFish:
\begin{itemize}
    \item \texttt{FFTW\_DataType} makes use of the highly optimised and popular  FFTW library (Fastest Fourier Transform in the West) \cite{FFTW05,FFTW97,FFTW_site}. This library has broad functionality and displays consistently good performance;
    \item \texttt{cuFFT\_DataType} makes use of the cuda language to leverage highly accelerated parallel execution on an NVIDIA graphical processing unit (GPU) \cite{cufft_site}. Although this library has far less functionality than FFTW, the speedups capable on a GPU unit allow for extremely fast execution on a desktop. 
\end{itemize}
Although these only represent two options, there are a range of other options depending on target device, compilation environment and intended solver type which may warrant the implementation of further \texttt{DataType} classes. A brief list of possible options include the Intel MKL FFT framework \cite{MKL_site}, the OpenCL FFT library \cite{OpenCLFFT_site}, or implementation in other languages such as Python FFT \cite{NumpyFFT_site}. Depending on the chosen data type, additional implementation may be necessary. 

\subsubsection{Solver Class}
The \texttt{Solver} parent class is derived directly from the user-defined \texttt{DataType} parent class and contains all of the functionality and interfacing methods for the FFT tool. By inheriting this in the solver implementations, an additional layer of abstraction is achieved. This class furthermore contains all of the virtual functions relevant to grid declaration and manipulation, solver option specification, preparation of the Green's functions and convolution execution. Four types of solver are derived from the \texttt{Solver} parent class: \texttt{1D\_Scalar}, \texttt{2D\_Scalar}, \texttt{3D\_Scalar} and \texttt{3D\_Vector}, named for the type of the source (and solution) data. The solvers relevant to specific BCs (see Section~\ref{sec:solvertypes}) are then derived from these classes. Within each solver class the following virtual methods must be declared and defined:
\begin{itemize}
    \item \texttt{FFT\_Data\_Setup}- Allocates data and FFT Transform types;
    \item \texttt{Specify\_Operator}- Specify desired spectral space operator types (Section \ref{sec:difops});
    \item \texttt{Forward\_Transform}- Form of the DFT from real to spectral space;
    \item \texttt{Backward\_Transform}- Form of the DFT from spectral to real space;
    \item \texttt{Convolution}- Form of the convolution (purely real, complex);
    \item \texttt{Specify\_Greens\_Function}- Calculates the Green's function on the grid for the convolution step.
\end{itemize}
Although these are declared and defined for the aforementioned solver types in SailFFish, the option exists for user defined solver classes with any additional operations necessary for a desired application.

\subsection{Generating and Executing a Solver}
A short overview is given here of the routines which are carried out to generate and execute a solver within SailFFish.

\subsubsection{Solver Initialization}
\label{sec:solverinit}
Numerous functions are called during the initialization of a solver. These are listed here:
\begin{enumerate}
    \item \texttt{X/Y/Z\_Grid\_Setup} Grid positions and statistics are specified.
    \item \texttt{Datatype\_Setup} Datatype-specific initializations are carried out. 
    \item \texttt{FFT\_Data\_Setup} FFT Data arrays are allocated and FFT plans are specified.
    \item \texttt{Specify\_Greens\_Function} Specification of either eigenvalues (bounded BCs) or Green's functions (unbounded BCs) arrays.
    \item \texttt{Prepare\_Dif\_Operators} Spectral-space differential operators are prepared (if appropriate).
\end{enumerate}
The specification of the Green's function is the most expensive process here, as many expensive function evaluations must be carried out. This process is parallelised with \texttt{OpenMP}. The computational expense of the initialization is not of significant consequence as this process is only carried out once at the beginning of a simulation. The solver may thereafter be executed arbitrarily many times. 

\subsubsection{Solver Execution}
\label{sec:solverexec}
The execution process is relatively straightforward for simplicity. The steps are listed chronologically:
\begin{enumerate}
    \item \texttt{Set\_Input} The values of $f$ on the grid are transferred into the necessary arrays for transforming.
    \item \texttt{Forward\_Transform} Forward FFTs are carried out on the data. 
    \item \texttt{Convolution} Convolution in the spectral space is carried out. This includes other procedures in the spectral space such as spectral differentiation.
    \item \texttt{Backward\_Transform} Backward (inverse) FFTs are carried out on the data. 
    \item \texttt{Get\_Output} The values of $\vec{\psi}$ on the grid are transferred into the necessary arrays for post-processing.
\end{enumerate}
The most expensive steps here are the forward and backward FFTs. An overview of solver timings is given in Section~\ref{sec:perf}

\section{Validation}
\label{sec:validation}
To demonstrate the convergence of the solvers a set of validation calculations have been carried out. Depending on the type of solver, a suitable source function has been chosen for which an analytical solution to the Poisson equation is known. It should be noted that the convergence achieved is a function of the input field $f$. The accuracy of the numerical solution depends on the number of continuous derivatives of the source function. It is therefore helpful for a convergence study to select a source function which displays $C^{\infty}$ smoothness. \newline
For all presented cases SailFFish has been configured to have double floating point accuracy. For all results shown, the convergence of the solver is demonstrated by showing the infinite norm of the error:
\begin{equation}
\label{eq:Einf}
E_{\infty} = \text{sup} \lbrace \psi(x,y,z) - \psi_{a}(x,y,z) \rbrace 
\text{ , }
\end{equation}
where $\psi_{a}$ represents the known analytical solution. This error norm is chosen as the $E_{\infty}$-norm bounds the $E_{2}$-norm. The absolute value of error is scaled by the lowest value of the solution $\psi_{a,min} = \text{inf}\lbrace\psi_{a}\rbrace $ on the grid. In order to make the error results comparable, all values of $E_{\infty}$ have been scaled such that they correspond to the grid solution with $\psi_{a,min}=1$. 

\subsection{Bounded Solvers: Homogeneous Boundary Conditions}
\begin{figure}
\centering
  \setlength\figureheight{3cm}
  \setlength\figurewidth{6.5cm}
  \vspace{6mm}
  \begin{tabular}{cc}
       \input{Plots/1D_CELL_CENTRE.tikz} &
       \input{Plots/1D_CELL_BDRY.tikz} \\
       \input{Plots/2D_CELL_CENTRE.tikz} &
       \input{Plots/2D_CELL_BDRY.tikz} \\
       \input{Plots/3D_CELL_CENTRE.tikz} &
       \input{Plots/3D_CELL_BDRY.tikz} \\
  \end{tabular}
\caption[]{Error of 1D (top row), 2D (middle row) and 3D (bottom row) solver types of staggered (left) and regular (right) grid configurations. The dashed line indicates quadratic decay.}
 \label{fig:BS}
\end{figure}
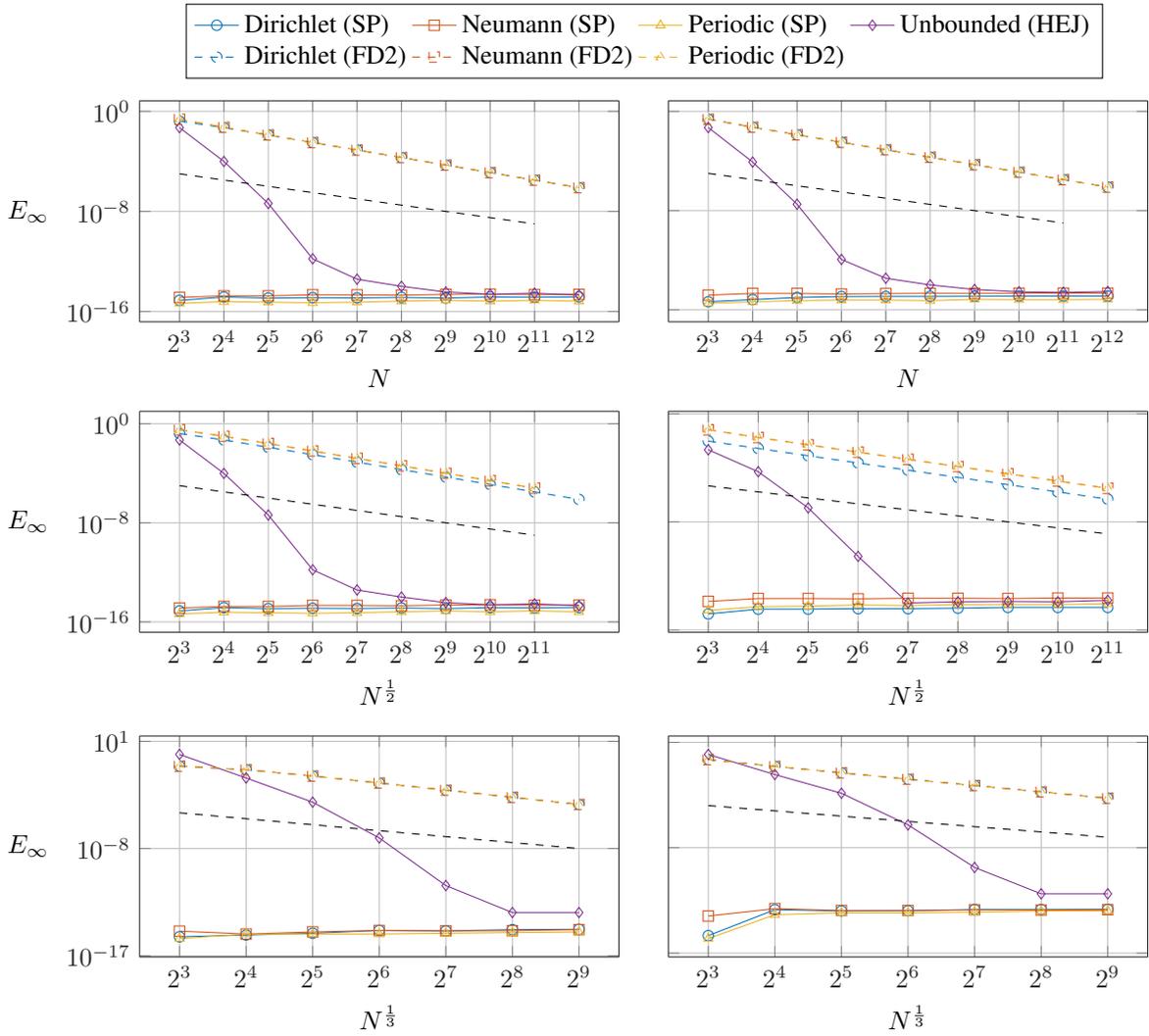

The first class of solvers investigated are the bounded solvers. The following 1D test functions have been used:
\begin{equation}
\label{eq:tfb}
S(x,L_x) = \sin \left(4\pi\frac{x}{L_x}\right) \text{, } \hspace{10mm}
C(x,L_x) = \cos \left(4\pi\frac{x}{L_x}\right) \text{, }
\end{equation}
where $L_x$ is the length of the domain in $x$ direction and the coordinate $x$ is taken to be relative to the lower bound of the $x$-domain ($x_l$). In higher dimensional cases, the product of this function in each direction is taken. Derivation of the analytical solution is trivial. For homogeneous Dirichlet BCs for example, the following function has been used:
\begin{equation}
\label{eq:tfb3D}
f_a(x,y,z) = S(x,L_x)\,S(y,L_y)\,S(z,L_z) 
\hspace{3mm} \longrightarrow \hspace{3mm}
\psi_a(x,y,z) = - (4\pi)^2 
\left( \frac{L_x+L_y+L_z}{L_x^2L_y^2L_z^2} \right) f_a(x,y,z)
\end{equation}
For homogeneous Neumann BCs, a similar expression to \eqref{eq:tfb} is used however with the cosine expression $C(x,L_x)$. The analytical solution corresponding to \eqref{eq:tfb3D} then follows easily. The error for the pseudo-spectral (PS) and second order finite difference (FD2) eigenvalues are shown in Fig.~\ref{fig:BS}. Results are shown for both regular and staggered grid configurations. As expected, application of the PS eigenvalues to the spectral source distribution yields spectral accuracy of the solution, as the error is seen to saturate at machine precision. Application of the FD2 type eigenvalues results in the expected quadratic convergence. For reference, the spectral 1D operator of Hejlesen is shown for a 1D unbounded problem \cite{Hejlesen_PhD}. It is observed that the expected behaviour is reproduced, independent of the grid configuration. The corresponding plots for the 2D and 3D cases are also shown in Fig.~\ref{fig:BS}. Practically exactly the same behaviour is observed in these cases as with the 1D case. It is seen that the 3D unbounded case does not quite reach spectral accuracy for the chosen grid size as the spatial resolution is lower than for the 1D and 2D cases (due to increased memory demand). The 3D-vector type solver results are not shown here as, by construction, these simply carry out the same procedures as the 3D scalar solver type. 
\subsection{Bounded Solvers: Inhomogeneous Boundary Conditions}
\begin{figure}[ht]
\centering
  \setlength\figureheight{3cm}
  \setlength\figurewidth{6.5cm}
  \vspace{15mm} 
  \begin{tabular}{cc}
       \input{Plots/2D_IH.tikz} &
       \input{Plots/3D_IH.tikz}
  \end{tabular}
\caption[]{Error of 2D (left) and 3D (right) solvers with inhomogeneous BCs for both staggered and regular grid configurations with FD2 type eigenvalues. The dashed line indicates quadratic decay.}
 \label{fig:IH}
\end{figure}
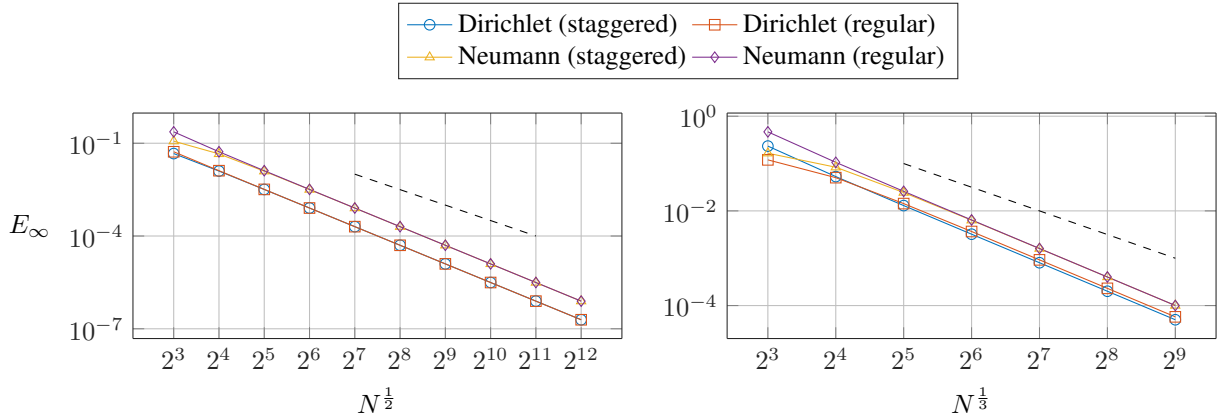

The option also exists within SailFFish to treat inhomogeneous BCs for the Dirichlet and Neumann bounded solvers. In this case, the form of the FD2 eigenvalue can be exploited to account for the desired boundary condition by simply modifying the interior boundary nodes values as described in Section~\ref{sec:coe}. This however excludes the use of the PS eigenvalue type.
The accuracy of these solution shall be investigated here. Inhomogeneous BCs are specified by applying the following input function:
\begin{equation}
\label{eq:tfih1D}
\psi_a(x,L_x) = a\left(1-\frac{x}{L_x}\right)^2 + b\left(\frac{x}{L_x}\right)^2
\hspace{2mm} \longrightarrow \hspace{2mm}
f_a(x,L_x) = 2.0\frac{a+b}{L_x^2} \text{.}
\end{equation}
This function represents a paraboloid with the values $\psi_a(0,L_x)=a$ and $\psi_b(L_x,L_x)=b$ on the lower and upper limits of the domain, respectively. In the 3D case, a function is simply added for each dimension. This results in a parabolic boundary distribution on each domain face, with corner values specified by the constants $a_x,b_x,a_y,\cdots$. This function furthermore has the advantage that the gradient function is easily calculated as:
\begin{equation}
\label{eq:tfih1Dgrad}
\frac{\partial \psi_a}{\partial x}(x,L_x) = 2a\left( \frac{x}{L_x} - 1 \right) + 2b\frac{x}{L_x} \text{ . }
\end{equation}
For test cases with Dirichlet BCs, \eqref{eq:tfih1D} is therefore applied to calculate the numerical value on the boundary. For test cases with Neumann BCs, \eqref{eq:tfih1Dgrad} is applied. As higher derivatives of this function vanish, even the FD2 type eigenvalue leads to spectral accuracy. For this reason, the test function is superimposed with the corresponding distribution from \eqref{eq:tfb}. For Dirichlet BCs, $S(\cdot)$ is applied and for Neumann BCs, $C(\cdot)$ is applied, such that the total BC on each face remains unchanged by this addition. The results for 2D and 3D cases are shown for a range of grid sizes in Fig.~\ref{fig:IH}. The 1D case is omitted for brevity, as the results are indistinguishable from the 2D and 3D cases. The error can be seen to decay quadratically as is expected for the FD2 eigenvalue type. The results here are seen to mirror those for homogeneous BCs in Fig.~\ref{fig:BS}. It should be noted that if all boundaries have Neumann-type BCs, the solution can only be determined up to a constant. An optimum solution which produces a solution with the lowest residual may be employed. This however is not implemented in SailFFish for simplicity. The results displayed in  Fig.~\ref{fig:IH} are therefore achieved by shifting the solution values such that $\text{sup}\lbrace\psi\rbrace=\text{sup}\lbrace\psi_a\rbrace$.

\subsection{Unbounded Solvers}
\label{sec:valunb}
As with the bounded solvers, it advantageous to apply a test function for which the boundary values are finite, however display $C^{\infty}$ smoothness. For the unbounded problems, a bump function has been applied. This is described as:
\begin{equation}
\label{eq:bump1D}
\psi_a(\zeta) = 
\begin{cases}
e^{-cd} & \zeta < 1 \\
0 & \zeta \geq 1
\end{cases}
\hspace{1mm} \rightarrow \hspace{1mm}
f_a(\zeta) = 
\begin{cases}
-\left( c^2\zeta^2d^4 -2cd^2 - 8c\zeta^2d^3\right)e^{-cd} & \zeta < 1 \\
0 & \zeta \geq 1
\end{cases} \text{ where: }
d = \frac{1}{1-\zeta^2} 
\end{equation}
and $c$ is a constant which was chosen for all cases to be 10. In the 1D case $\zeta$ is chosen to be the absolute distance from the origin. In the 2D case $\zeta$ is taken to be the radius from the origin in the $x$-$y$ plane with: $\zeta=\sqrt{x^2+y^2}$ and the vorticity vector is aligned with the $z$ axis. In the 3D case a vortex ring of radius $r_0$ in the $y$-$z$ plane is prescribed. The vorticity distribution in any plane $\theta = \tan^{-1}(z/y) =$ const. is described by \eqref{eq:bump1D} where $\zeta$ is the distance to the vortex ring coordinate in that plane: $\zeta = \sqrt{(r-r_0)^2 + z^2}$ and the vorticity vector is aligned with the tangent to the vortex ring at that position. Due to the additional radial terms of the Laplacian in cylindrical coordinates, the expressions for $f_a$ become more complex in 2D and 3D cases. These are omitted here for brevity, however the expressions can be found in the corresponding unbounded test function files in the library  \href{https://github.com/ZeppSav/SailFFish}{repository}. \newline 
In the case of unbounded BCs, the Green's function is first calculated in real space, before being forward transformed to the spectral space. This procedure is essentially independent of whether staggered or regular grid types are applied (again, this is true for the currently implemented cases of uniform BC types, the situation differs for mixed BCs \cite{Caprace_2021}). For this reason, in the following results only staggered results are shown for brevity. The results are identical for regular grid types. As described in Section~\ref{sec:unbounded}, a range of kernels have been implemented for unbounded BCs including Gaussian approximations from order 2-10 and the spectral kernel representation, all attributable to Hejlesen \cite{Hejlesen_PhD}. For the 1D case only the spectral-type kernel is employed. For brevity, 1D results are not shown here, however the spectral convergence can be seen for the 1D cases in Fig.~\ref{fig:BS}.
\begin{figure}[ht]
\centering
  \setlength\figureheight{3cm}
  \setlength\figurewidth{6.5cm}
  \vspace{10mm} 
  \begin{tabular}{cc}
       \input{Plots/2D_UB.tikz} &
       \input{Plots/3D_UB.tikz}
  \end{tabular}
\caption[]{Error of 2D (left) and 3D (right) solvers with unbounded BCs for a range of unbounded kernel types.}
 \label{fig:UB}
\end{figure}
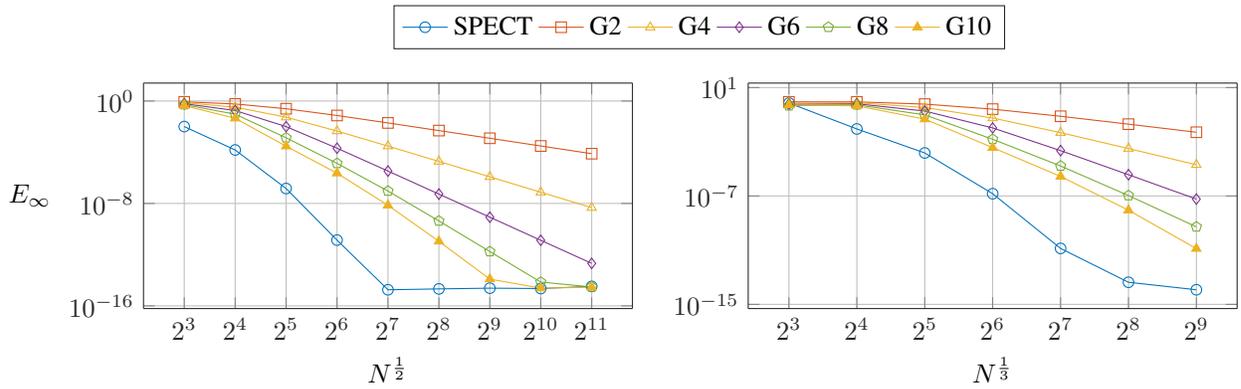
The results for the bump function in 2D and 3D are given in Fig.~\ref{fig:UB}. It can be seen that the convergence order of the different kernels follows precisely the behaviour expected for the given kernel order. The grid size in the 3D extends only up to $[512,512,512]$ due to the memory overhead of the Hockney-Eastwood method and the available memory on the device employed. Despite this, the convergence behaviour is precisely as expected.
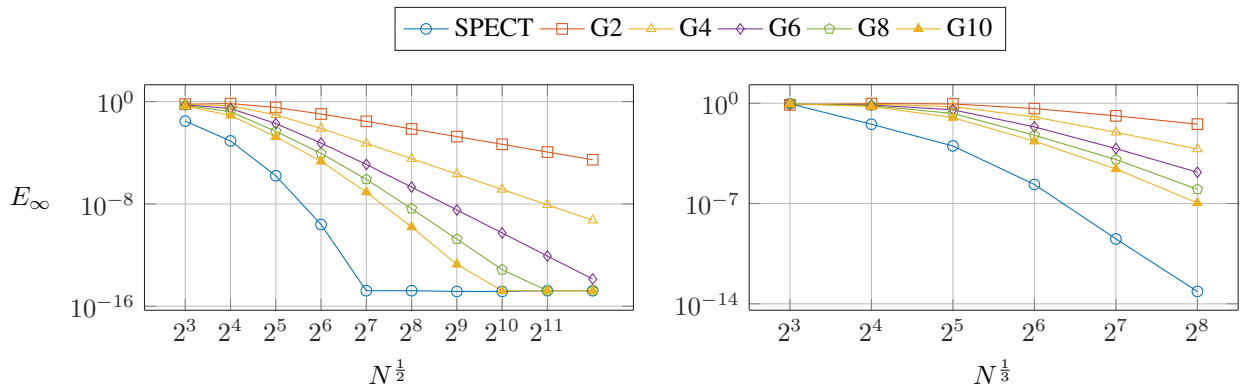
\begin{figure}[ht]
\centering
  \setlength\figureheight{3cm}
  \setlength\figurewidth{6.5cm}
  \vspace{10mm}
  \begin{tabular}{cc}
       \input{Plots/2D_UB_Grad.tikz} &
       \input{Plots/3D_UB_Grad.tikz}
  \end{tabular}
\caption[]{Error of 2D (left) and 3D (right) solvers with unbounded BCs with the grad (2D) and curl (3D) differential operator for a range of unbounded kernel types.}
 \label{fig:UBGrad}
\end{figure}
The correct functioning of the differential operators has been tested by applying to the distributions described previously, however for the solution of the curl Poisson equation:
\begin{equation}
\label{eq:poisscurl}
\nabla^2\vec{\psi} = \nabla\times\vec{f} \text{ .}
\end{equation}
In this case, the same solution procedure is carried out as for the standard unbounded solver, however after convolution spatial differentiation is carried out on the data in the spectral space via \eqref{eq:ft_sd} in order to extract the derivatives. This form of the equation may be applied, for example, when extracting the velocity field from the vorticity field in fluid dynamics \cite{Chatelain_2010, Hejlesen_PhD}. In this case, analytical expressions also exist for the 2D and 3D bump functions and are again omitted here for brevity. The 2D curl operator is practically equivalent to the application of the gradient operator. As compared to the previous unbounded tests cases, the application of these gradient terms results in multiple output fields, corresponding to the velocity components of the extracted field. In this case, $E_{\infty}$ is applied to all velocity components to extract the maximum absolute error. The results for the application of spectral differentiation to the bump function in 2D and 3D are shown in Fig.~\ref{fig:UBGrad}. The error is again seen to scale with the same behaviour as for the direct Poisson equation for the bump function. This demonstrates the efficacy of the spectral differentiation implemented with SailFFish.

\section{Performance}
\label{sec:perf}
The performance of the library is now investigated. An inspection of timings is provided, scaling of the solver is discussed and finally a breakdown of the solver timings is given for a range of simulation setups. As stated in Section~\ref{sec:solverinit}, solver initialization only occurs once and thus the expense of this step is of less consequence than the solver timings. For the following discussion, steps 2-4 of the solver execution (see Section~\ref{sec:solverexec}) are considered for the timings, as the input and output steps are generally of low computational expense and are also likely to be very application-specific. These function serve only as an access point to the underlying data arrays. \newline
The simulations carried out here were done so on a desktop computer with an Intel(R) 12$^{th}$ Gen Core$^{\text{TM}}$ i7-12700KF processor with a clock speed of 3.60 GHz, 20 available threads and 64 GB of installed RAM. For simulations utilizing the \texttt{cuFFT\_DataType}, execution was carried out on an NVIDIA GeForce RTW 3070 Ti graphics card with 8GB of on-device memory. All simulations were carried out with single floating point accuracy. All data points shown were gathered by averaging the execution times over ten simulation runs.

\subsection{Execution Time}
\begin{figure}[ht]
\centering
  \setlength\figureheight{4cm}
  \setlength\figurewidth{8cm}
  \input{Plots/Perf1.tikz}
\caption[]{Calculation times in microseconds for a range of problem sizes using both FFTW and CUDA datatypes. The solver type is a bounded 3D solver with periodic BCs.}
 \label{fig:Perf1}
\end{figure}
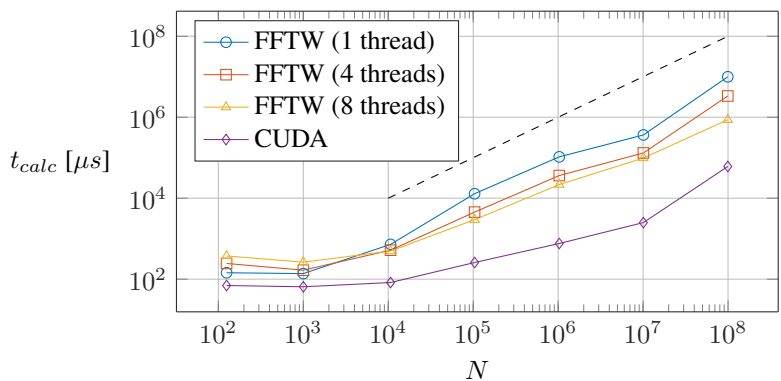

A plot of solver wall clock timings is given in Fig.~\ref{fig:Perf1}. As mentioned previously, these timings are the sum total of forward transformation, convolution, and backward transformation. Results are shown for both \texttt{FFTW\_DataType} and \texttt{cuFFT\_DataType} base classes. It is observed that for lower $N$, activating multi-threading within FFTW has a negative effect due to the additional overhead associated with the use \texttt{OpenMP} and the comparatively small problem size. The advantages can however be seen for larger $N$. With the use of eight threads, an improvement of an order of magnitude is observed as compared to single-thread execution. It is also observed that the use of the GPU for the calculation drastically improves performance. As compared to CPU operation an improvement in calculation time of between one and two orders of magnitude can be observed.  
As described in Section~\ref{sec:framework}, it is desired to minimise solver communication overheads. Ideally, memory allocation for the FFT data arrays should therefore occur directly on the GPU and transfers between CPU and GPU should be avoided. This limits directly the allowable problem size to the available memory on the GPU. These limits may be extended with the use of half-floating point precision, provided the GPU architecture allows for this. 

\subsection{Scalability}
As SailFFish is configured for shared-memory execution, strong scaling shall be discussed. The three components of the execution time are all easily multi-threaded with \texttt{OpenMP}. The convolution procedure is nothing more than a vector-vector multiplication, as are any of the spectral differentiation routines. The scaling here is therefore device-relevant and constrained by the same overhead limitation as described in the previous section. For the \texttt{FFTW\_DataType}, the forward and backward transform steps are nothing more than a forward and backward execution within FFTW (e.g. \texttt{fftw\_execute\_dft} function) and are therefore subject to the multi-threaded implementation of FFTW. It was difficult to attain precisely linear scaling as there appeared to be a significant overhead between the cases of a single and multiple threads. For some problem sizes linear scaling was observed between 4-28 threads. For other cases, a discontinuous jump in execution time resulted above a given number of threads. Further investigations are required here to establish robust scaling behaviour. In most cases however, as observed in Fig.~\ref{fig:Perf1}, the use of multiple threads greatly improves performance. Similar investigations are not possible for the \texttt{cuFFT\_DataType}, as specifying the number of active GPU threads cannot be easily facilitated.

\subsection{Breakdown of Execution Steps}
\begin{figure}[ht]
\centering
  \setlength\figureheight{6cm}
  \setlength\figurewidth{15cm}
  \input{Plots/Bar.tikz}
\caption[]{Proportion of calculation time for a range of problem sizes, datatype and differential operators. For solver parameters, see Table~\ref{tab:PropTrials}.}
 \label{fig:Fracs}
\end{figure}
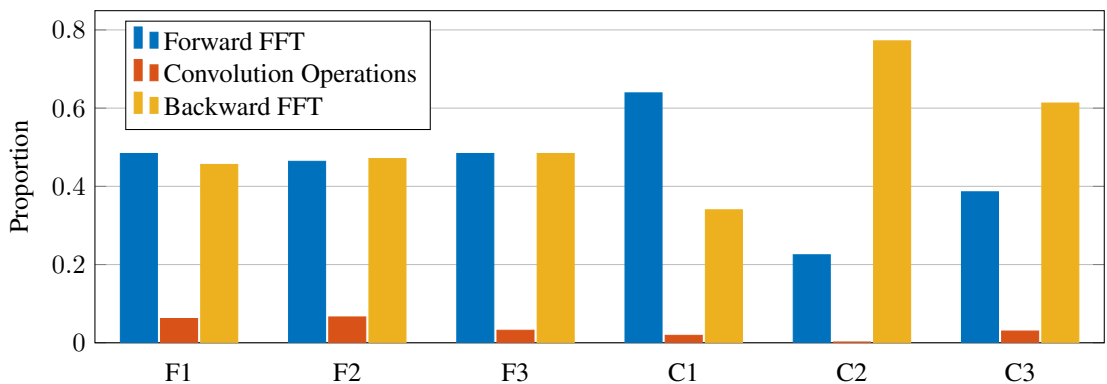

In order to investigate the relative computational expense of the different processes of the execution, these are broken down and relative execution times are compared. A range of problem sizes and DataType were investigated, these are summarised in Table~\ref{tab:PropTrials}. Two cases were considered with the curl operator, as this is particularly relevant to the solution of e.g. \eqref{eq:poisscurl}. For the (F-type) simulations carried out here, eight threads have been applied concurrently for calculation. Similar behaviour was however observed that for any choice of number of concurrent threads.
\begin{table}[ht]
 \caption{Simulation Parameters for Fig.~\ref{fig:Fracs}}
  \centering
  \begin{tabular}{|ccccc|}
    \toprule
    Simulation & Datatype & $N$ Nodes & BC Type & Operator \\ \hline \hline 
    F1 & FFTW & $64^3$  & Periodic  & --- \\
    F2 & FFTW & $128^3$ & Unbounded & CURL \\
    F3 & FFTW & $512^3$ & Periodic  & --- \\
    C1 & CUDA & $64^3$  & Periodic  & --- \\
    C2 & CUDA & $128^3$ & Unbounded & CURL \\
    C3 & CUDA & $512^3$ & Periodic  & --- \\
    \bottomrule
    \end{tabular}
    \label{tab:PropTrials}
\end{table}
The results for these cases are shown in Fig.~\ref{fig:Fracs}. A number of observations can be made. Convolution procedures generally consume only a small portion of the total execution time, including cases where a differential operator is being applied in the spectral space. This portion increases for smaller problem sizes. The convolution times of the \texttt{cuFFT\_DataType} are generally smaller. This is likely due to the (adapted) use of a highly optimised matrix-vector multiplication algorithm for convolution procedures. It is also observed that an asymmetry occurs for the \texttt{cuFFT\_DataType}. The causes for this require further investigation.  






\section{Conclusions \& Outlook}
\label{sec:conc}
SailFFish is a lightweight, parallelised library for the solution of the Poisson equation on a rectangular grid with regular grid spacing. The solver is constructed with the goals of simplicity and adaptability and is optimised for execution within shared-memory applications. The solver is capable of handing bounded Poisson problems whereby Dirichlet, Neumann or periodic boundary conditions are specified on the domain boundary. Spectral eigenvalue specification allows for the solution of the Poisson equation with spectral accuracy for homogeneous boundary conditions. The use of the second-order finite difference eigenvalues additionally allows for the treatment of arbitrary inhomogeneous boundary conditions of  Dirichlet or Neumann type. Unbounded boundary conditions can also be treated within SailFFish through application of the Hockney-Eastwood method of domain doubling. In this case a range of Green's function are available. Currently only uniform boundary conditions are available, however it is desired to implement mixed boundary conditions in the near future. It is hoped that in the future an optimised James-Lacker algorithm may be applied to reduce memory overhead for the case of unbounded boundary conditions. All solver types can be executed with either regular or staggered grid configurations. All solver configuration were validated against analytical solutions to the Poisson equation and were found to converge at the expected rate. \\ \\
A description of the solver framework was given to provide insight for future adaptation of the solver. The solver is currently implemented with two FFT frameworks for transforms and operations within the spectral space:
\begin{itemize}
    \item \texttt{FFTW\_DataType}, whereby the FFTW library is leveraged for optimised calculation on a CPU; and 
    \item \texttt{cuFFT\_DataType}, whereby the cuFFT library is leveraged for optimised calculation on a GPU.
\end{itemize}
It is hope that within an open-source context, future \texttt{DataType} formats will be implemented as new, optimised libraries become available. The development of SailFFish is a first step towards the implementation of a GPU-accelerated solver for the vorticity transport equation.

\section*{Acknowledgments}
The author would like to acknowledge Denis-Gabriel Caprace for many helpful, interesting and enjoyable conversations on the topic of fast Poisson solvers and their application. 
\bibliographystyle{unsrt}  
\bibliography{references}

\end{document}

%% file: Plots/1D_CELL_CENTRE.tikz
%
%
%
\definecolor{mycolor1}{rgb}{0,0.447,0.741}%
\definecolor{mycolor2}{rgb}{0.85,0.325,0.098}%
\definecolor{mycolor3}{rgb}{0.929,0.694,0.125}%
\definecolor{mycolor4}{rgb}{0.494,0.184,0.556}%
\definecolor{mycolor5}{rgb}{0.466,0.674,0.188}%
\begin{tikzpicture}

\begin{axis}[%
width=\figurewidth,
height=\figureheight,
scale only axis,
separate axis lines,
every outer x axis line/.append style={darkgray!60!black},
every x tick label/.append style={font=\color{darkgray!60!black}},
xmode=log,
xtick={8,16,32,64,128,256,512,1024,2048,4096},
xticklabels={$2^3$,$2^4$,$2^5$,$2^6$,$2^7$,$2^8$,$2^9$,$2^{10}$,$2^{11}$,$2^{12}$},
xlabel={$N$},
xmajorgrids,
every outer y axis line/.append style={darkgray!60!black},
every y tick label/.append style={font=\color{darkgray!60!black}},
ymode=log,
yminorticks=true,
ylabel={$E_{\infty}$},
ylabel near ticks,
ylabel style={rotate=-90},
ymajorgrids,
yminorgrids,
legend style={
draw=darkgray!60!black,
fill=white,
at={(1.1,0.5)},anchor=west,		
legend cell align=left				
}
]





\addplot [color=mycolor1,mark=o]
table[x index=0,y index=1]{Plots/1D_CELL_CENTRE_PS.dat};

\addplot [color=mycolor2,mark=square]
table[x index=0,y index=2]{Plots/1D_CELL_CENTRE_PS.dat};

\addplot [color=mycolor3,mark=triangle]
table[x index=0,y index=3]{Plots/1D_CELL_CENTRE_PS.dat};

\addplot [color=mycolor4,mark=diamond]
table[x index=0,y index=4]{Plots/1D_CELL_CENTRE_PS.dat};

\addplot [color=mycolor1,mark=o, dashed]
table[x index=0,y index=1]{Plots/1D_CELL_CENTRE_FD2.dat};

\addplot [color=mycolor2,mark=square, dashed]
table[x index=0,y index=2]{Plots/1D_CELL_CENTRE_FD2.dat};

\addplot [color=mycolor3,mark=triangle, dashed]
table[x index=0,y index=3]{Plots/1D_CELL_CENTRE_FD2.dat};

\addplot [color=black,dashed]
table[row sep=crcr]{
8 1e-5\\
2048 1e-9\\
};


\end{axis}
\end{tikzpicture}%

%% file: Plots/1D_CELL_BDRY.tikz
%
%
%
\definecolor{mycolor1}{rgb}{0,0.447,0.741}%
\definecolor{mycolor2}{rgb}{0.85,0.325,0.098}%
\definecolor{mycolor3}{rgb}{0.929,0.694,0.125}%
\definecolor{mycolor4}{rgb}{0.494,0.184,0.556}%
\definecolor{mycolor5}{rgb}{0.466,0.674,0.188}%
\begin{tikzpicture}

\begin{axis}[%
width=\figurewidth,
height=\figureheight,
scale only axis,
separate axis lines,
every outer x axis line/.append style={darkgray!60!black},
every x tick label/.append style={font=\color{darkgray!60!black}},
xmode=log,
xtick={8,16,32,64,128,256,512,1024,2048,4096},
xticklabels={$2^3$,$2^4$,$2^5$,$2^6$,$2^7$,$2^8$,$2^9$,$2^{10}$,$2^{11}$,$2^{12}$},
xlabel={$N$},
xmajorgrids,
every outer y axis line/.append style={darkgray!60!black},
every y tick label/.append style={font=\color{darkgray!60!black}},
ymode=log,
yminorticks=true,
yticklabels={,,},
ymajorgrids,
yminorgrids,
legend style={
fill=white,
legend columns=4,
at={(-0.05,1.1)},anchor=south,		
overlay,
legend cell align=left				
}
]





\addplot [color=mycolor1,mark=o]
table[x index=0,y index=1]{Plots/1D_CELL_BDRY_PS.dat};
\addlegendentry{Dirichlet (SP)};

\addplot [color=mycolor2,mark=square]
table[x index=0,y index=2]{Plots/1D_CELL_BDRY_PS.dat};
\addlegendentry{Neumann (SP)};

\addplot [color=mycolor3,mark=triangle]
table[x index=0,y index=3]{Plots/1D_CELL_BDRY_PS.dat};
\addlegendentry{Periodic (SP)};

\addplot [color=mycolor4,mark=diamond]
table[x index=0,y index=4]{Plots/1D_CELL_BDRY_PS.dat};
\addlegendentry{Unbounded (HEJ)};

\addplot [color=mycolor1,mark=o, dashed]
table[x index=0,y index=1]{Plots/1D_CELL_BDRY_FD2.dat};
\addlegendentry{Dirichlet (FD2)};

\addplot [color=mycolor2,mark=square, dashed]
table[x index=0,y index=2]{Plots/1D_CELL_BDRY_FD2.dat};
\addlegendentry{Neumann (FD2)};

\addplot [color=mycolor3,mark=triangle, dashed]
table[x index=0,y index=3]{Plots/1D_CELL_BDRY_FD2.dat};
\addlegendentry{Periodic (FD2)};

\addplot [color=black,dashed]
table[row sep=crcr]{
8 1e-5\\
2048 1e-9\\
};


\end{axis}
\end{tikzpicture}%

%% file: Plots/2D_CELL_CENTRE.tikz
%
%
%
\definecolor{mycolor1}{rgb}{0,0.447,0.741}%
\definecolor{mycolor2}{rgb}{0.85,0.325,0.098}%
\definecolor{mycolor3}{rgb}{0.929,0.694,0.125}%
\definecolor{mycolor4}{rgb}{0.494,0.184,0.556}%
\definecolor{mycolor5}{rgb}{0.466,0.674,0.188}%
\begin{tikzpicture}

\begin{axis}[%
width=\figurewidth,
height=\figureheight,
scale only axis,
separate axis lines,
every outer x axis line/.append style={darkgray!60!black},
every x tick label/.append style={font=\color{darkgray!60!black}},
xmode=log,
xtick={8,16,32,64,128,256,512,1024,2048},
xticklabels={$2^3$,$2^4$,$2^5$,$2^6$,$2^7$,$2^8$,$2^9$,$2^{10}$,$2^{11}$},
xlabel={$N^{\frac{1}{2}}$},
xmajorgrids,
every outer y axis line/.append style={darkgray!60!black},
every y tick label/.append style={font=\color{darkgray!60!black}},
ymode=log,
yminorticks=true,
ylabel={$E_{\infty}$},
ylabel near ticks,
ylabel style={rotate=-90},
ymajorgrids,
yminorgrids,
legend style={
draw=darkgray!60!black,
fill=white,
at={(1.1,0.5)},anchor=west,		
legend cell align=left				
}
]





\addplot [color=mycolor1,mark=o]
table[x index=0,y index=1]{Plots/1D_CELL_CENTRE_PS.dat};

\addplot [color=mycolor2,mark=square]
table[x index=0,y index=2]{Plots/1D_CELL_CENTRE_PS.dat};

\addplot [color=mycolor3,mark=triangle]
table[x index=0,y index=3]{Plots/1D_CELL_CENTRE_PS.dat};

\addplot [color=mycolor4,mark=diamond]
table[x index=0,y index=4]{Plots/1D_CELL_CENTRE_PS.dat};

\addplot [color=mycolor1,mark=o, dashed]
table[x index=0,y index=1]{Plots/1D_CELL_CENTRE_FD2.dat};

\addplot [color=mycolor2,mark=square, dashed]
table[x index=0,y index=2]{Plots/2D_CELL_CENTRE_FD2.dat};

\addplot [color=mycolor3,mark=triangle, dashed]
table[x index=0,y index=3]{Plots/2D_CELL_CENTRE_FD2.dat};

\addplot [color=black,dashed]
table[row sep=crcr]{
8 1e-5\\
2048 1e-9\\
};


\end{axis}
\end{tikzpicture}%

%% file: Plots/2D_CELL_BDRY.tikz
%
%
%
\definecolor{mycolor1}{rgb}{0,0.447,0.741}%
\definecolor{mycolor2}{rgb}{0.85,0.325,0.098}%
\definecolor{mycolor3}{rgb}{0.929,0.694,0.125}%
\definecolor{mycolor4}{rgb}{0.494,0.184,0.556}%
\definecolor{mycolor5}{rgb}{0.466,0.674,0.188}%
\begin{tikzpicture}

\begin{axis}[%
width=\figurewidth,
height=\figureheight,
scale only axis,
separate axis lines,
every outer x axis line/.append style={darkgray!60!black},
every x tick label/.append style={font=\color{darkgray!60!black}},
xmode=log,
xtick={8,16,32,64,128,256,512,1024,2048},
xticklabels={$2^3$,$2^4$,$2^5$,$2^6$,$2^7$,$2^8$,$2^9$,$2^{10}$,$2^{11}$},
xlabel={$N^{\frac{1}{2}}$},
xmajorgrids,
every outer y axis line/.append style={darkgray!60!black},
every y tick label/.append style={font=\color{darkgray!60!black}},
ymode=log,
yminorticks=true,
yticklabels={,,},
ymajorgrids,
yminorgrids,
]





\addplot [color=mycolor1,mark=o]
table[x index=0,y index=1]{Plots/2D_CELL_BDRY_PS.dat};

\addplot [color=mycolor2,mark=square]
table[x index=0,y index=2]{Plots/2D_CELL_BDRY_PS.dat};

\addplot [color=mycolor3,mark=triangle]
table[x index=0,y index=3]{Plots/2D_CELL_BDRY_PS.dat};

\addplot [color=mycolor4,mark=diamond]
table[x index=0,y index=4]{Plots/2D_CELL_BDRY_PS.dat};

\addplot [color=mycolor1,mark=o, dashed]
table[x index=0,y index=1]{Plots/2D_CELL_BDRY_FD2.dat};

\addplot [color=mycolor2,mark=square, dashed]
table[x index=0,y index=2]{Plots/2D_CELL_BDRY_FD2.dat};

\addplot [color=mycolor3,mark=triangle, dashed]
table[x index=0,y index=3]{Plots/2D_CELL_BDRY_FD2.dat};

\addplot [color=black,dashed]
table[row sep=crcr]{
8 1e-5\\
2048 1e-9\\
};


\end{axis}
\end{tikzpicture}%

%% file: Plots/3D_CELL_CENTRE.tikz
%
%
%
\definecolor{mycolor1}{rgb}{0,0.447,0.741}%
\definecolor{mycolor2}{rgb}{0.85,0.325,0.098}%
\definecolor{mycolor3}{rgb}{0.929,0.694,0.125}%
\definecolor{mycolor4}{rgb}{0.494,0.184,0.556}%
\definecolor{mycolor5}{rgb}{0.466,0.674,0.188}%
\begin{tikzpicture}

\begin{axis}[%
width=\figurewidth,
height=\figureheight,
scale only axis,
separate axis lines,
every outer x axis line/.append style={darkgray!60!black},
every x tick label/.append style={font=\color{darkgray!60!black}},
xmode=log,
xtick={8,16,32,64,128,256,512},
xticklabels={$2^3$,$2^4$,$2^5$,$2^6$,$2^7$,$2^8$,$2^9$},
xlabel={$N^{\frac{1}{3}}$},
xmajorgrids,
every outer y axis line/.append style={darkgray!60!black},
every y tick label/.append style={font=\color{darkgray!60!black}},
ymode=log,
yminorticks=true,
ylabel={$E_{\infty}$},
ylabel near ticks,
ylabel style={rotate=-90},
ymajorgrids,
yminorgrids,
legend style={
draw=darkgray!60!black,
fill=white,
at={(1.1,0.5)},anchor=west,		
legend cell align=left				
}
]





\addplot [color=mycolor1,mark=o]
table[x index=0,y index=1]{Plots/3D_CELL_CENTRE_PS.dat};

\addplot [color=mycolor2,mark=square]
table[x index=0,y index=2]{Plots/3D_CELL_CENTRE_PS.dat};

\addplot [color=mycolor3,mark=triangle]
table[x index=0,y index=3]{Plots/3D_CELL_CENTRE_PS.dat};

\addplot [color=mycolor4,mark=diamond]
table[x index=0,y index=4]{Plots/3D_CELL_CENTRE_PS.dat};

\addplot [color=mycolor1,mark=o, dashed]
table[x index=0,y index=1]{Plots/3D_CELL_CENTRE_FD2.dat};

\addplot [color=mycolor2,mark=square, dashed]
table[x index=0,y index=2]{Plots/3D_CELL_CENTRE_FD2.dat};

\addplot [color=mycolor3,mark=triangle, dashed]
table[x index=0,y index=3]{Plots/3D_CELL_CENTRE_FD2.dat};

\addplot [color=black,dashed]
table[row sep=crcr]{
8 1e-5\\
512 1e-8\\
};


\end{axis}
\end{tikzpicture}%

%% file: Plots/3D_CELL_BDRY.tikz
%
%
%
\definecolor{mycolor1}{rgb}{0,0.447,0.741}%
\definecolor{mycolor2}{rgb}{0.85,0.325,0.098}%
\definecolor{mycolor3}{rgb}{0.929,0.694,0.125}%
\definecolor{mycolor4}{rgb}{0.494,0.184,0.556}%
\definecolor{mycolor5}{rgb}{0.466,0.674,0.188}%
\begin{tikzpicture}

\begin{axis}[%
width=\figurewidth,
height=\figureheight,
scale only axis,
separate axis lines,
every outer x axis line/.append style={darkgray!60!black},
every x tick label/.append style={font=\color{darkgray!60!black}},
xmode=log,
xtick={8,16,32,64,128,256,512},
xticklabels={$2^3$,$2^4$,$2^5$,$2^6$,$2^7$,$2^8$,$2^9$},
xlabel={$N^{\frac{1}{3}}$},
xmajorgrids,
every outer y axis line/.append style={darkgray!60!black},
every y tick label/.append style={font=\color{darkgray!60!black}},
ymode=log,
yminorticks=true,
yticklabels={,,},
ymajorgrids,
yminorgrids,
]





\addplot [color=mycolor1,mark=o]
table[x index=0,y index=1]{Plots/3D_CELL_BDRY_PS.dat};

\addplot [color=mycolor2,mark=square]
table[x index=0,y index=2]{Plots/3D_CELL_BDRY_PS.dat};

\addplot [color=mycolor3,mark=triangle]
table[x index=0,y index=3]{Plots/3D_CELL_BDRY_PS.dat};

\addplot [color=mycolor4,mark=diamond]
table[x index=0,y index=4]{Plots/3D_CELL_BDRY_PS.dat};

\addplot [color=mycolor1,mark=o, dashed]
table[x index=0,y index=1]{Plots/3D_CELL_BDRY_FD2.dat};

\addplot [color=mycolor2,mark=square, dashed]
table[x index=0,y index=2]{Plots/3D_CELL_BDRY_FD2.dat};

\addplot [color=mycolor3,mark=triangle, dashed]
table[x index=0,y index=3]{Plots/3D_CELL_BDRY_FD2.dat};

\addplot [color=black,dashed]
table[row sep=crcr]{
8 1e-5\\
512 1e-8\\
};


\end{axis}
\end{tikzpicture}%

%% file: Plots/2D_IH.tikz
%
%
%
\definecolor{mycolor1}{rgb}{0,0.447,0.741}%
\definecolor{mycolor2}{rgb}{0.85,0.325,0.098}%
\definecolor{mycolor3}{rgb}{0.929,0.694,0.125}%
\definecolor{mycolor4}{rgb}{0.494,0.184,0.556}%
\definecolor{mycolor5}{rgb}{0.466,0.674,0.188}%
\begin{tikzpicture}

\begin{axis}[%
width=\figurewidth,
height=\figureheight,
scale only axis,
separate axis lines,
every outer x axis line/.append style={darkgray!60!black},
every x tick label/.append style={font=\color{darkgray!60!black}},
xmode=log,
xtick={8,16,32,64,128,256,512,1024,2048,4096},
xticklabels={$2^3$,$2^4$,$2^5$,$2^6$,$2^7$,$2^8$,$2^9$,$2^{10}$,$2^{11}$,$2^{12}$},
xlabel={$N^{\frac{1}{2}}$},
xmajorgrids,
every outer y axis line/.append style={darkgray!60!black},
every y tick label/.append style={font=\color{darkgray!60!black}},
ymode=log,
yminorticks=true,
ylabel={$E_{\infty}$},
ylabel near ticks,
ylabel style={rotate=-90},
ymajorgrids,
yminorgrids,
legend style={
draw=darkgray!60!black,
fill=white,
at={(1.1,0.5)},anchor=west,		
legend cell align=left				
}
]





\addplot [color=mycolor1,mark=o]
table[x index=0,y index=1]{Plots/2D_IH.dat};

\addplot [color=mycolor2,mark=square]
table[x index=0,y index=2]{Plots/2D_IH.dat};

\addplot [color=mycolor3,mark=triangle]
table[x index=0,y index=3]{Plots/2D_IH.dat};

\addplot [color=mycolor4,mark=diamond]
table[x index=0,y index=4]{Plots/2D_IH.dat};

\addplot [color=black,dashed]
table[row sep=crcr]{
128 1e-2\\
2048 1e-4\\
};


\end{axis}
\end{tikzpicture}%

%% file: Plots/3D_IH.tikz
%
%
%
\definecolor{mycolor1}{rgb}{0,0.447,0.741}%
\definecolor{mycolor2}{rgb}{0.85,0.325,0.098}%
\definecolor{mycolor3}{rgb}{0.929,0.694,0.125}%
\definecolor{mycolor4}{rgb}{0.494,0.184,0.556}%
\definecolor{mycolor5}{rgb}{0.466,0.674,0.188}%
\begin{tikzpicture}

\begin{axis}[%
width=\figurewidth,
height=\figureheight,
scale only axis,
separate axis lines,
every outer x axis line/.append style={darkgray!60!black},
every x tick label/.append style={font=\color{darkgray!60!black}},
xmode=log,
xtick={8,16,32,64,128,256,512},
xticklabels={$2^3$,$2^4$,$2^5$,$2^6$,$2^7$,$2^8$,$2^9$},
xlabel={$N^{\frac{1}{3}}$},
xmajorgrids,
every outer y axis line/.append style={darkgray!60!black},
every y tick label/.append style={font=\color{darkgray!60!black}},
ymode=log,
yminorticks=true,
ymajorgrids,
yminorgrids,
legend style={
draw=darkgray!60!black,
fill=white,
legend columns=2,
overlay,
at={(-0.1,1.15)},anchor=south,		
legend cell align=left				
}
]





\addplot [color=mycolor1,mark=o]
table[x index=0,y index=1]{Plots/3D_IH.dat};
\addlegendentry{Dirichlet (staggered)};

\addplot [color=mycolor2,mark=square]
table[x index=0,y index=2]{Plots/3D_IH.dat};
\addlegendentry{Dirichlet (regular)};

\addplot [color=mycolor3,mark=triangle]
table[x index=0,y index=3]{Plots/3D_IH.dat};
\addlegendentry{Neumann (staggered)};

\addplot [color=mycolor4,mark=diamond]
table[x index=0,y index=4]{Plots/3D_IH.dat};
\addlegendentry{Neumann (regular)};

\addplot [color=black,dashed]
table[row sep=crcr]{
32 1e-1\\
512 1e-3\\
};


\end{axis}
\end{tikzpicture}%

%% file: Plots/2D_UB.tikz
%
%
%
\definecolor{mycolor1}{rgb}{0,0.447,0.741}%
\definecolor{mycolor2}{rgb}{0.85,0.325,0.098}%
\definecolor{mycolor3}{rgb}{0.929,0.694,0.125}%
\definecolor{mycolor4}{rgb}{0.494,0.184,0.556}%
\definecolor{mycolor5}{rgb}{0.466,0.674,0.188}%
\begin{tikzpicture}

\begin{axis}[%
width=\figurewidth,
height=\figureheight,
scale only axis,
separate axis lines,
every outer x axis line/.append style={darkgray!60!black},
every x tick label/.append style={font=\color{darkgray!60!black}},
xmode=log,
xtick={8,16,32,64,128,256,512,1024,2048},
xticklabels={$2^3$,$2^4$,$2^5$,$2^6$,$2^7$,$2^8$,$2^9$,$2^{10}$,$2^{11}$},
xlabel={$N^{\frac{1}{2}}$},
xmajorgrids,
every outer y axis line/.append style={darkgray!60!black},
every y tick label/.append style={font=\color{darkgray!60!black}},
ymode=log,
yminorticks=true,
ylabel={$E_{\infty}$},
ylabel near ticks,
ylabel style={rotate=-90},
ymajorgrids,
yminorgrids,
legend style={
draw=darkgray!60!black,
fill=white,
at={(1.1,0.5)},anchor=west,		
legend cell align=left				
}
]





\addplot [color=mycolor1,mark=o]
table[x index=0,y index=1]{Plots/2D_UB.dat};

\addplot [color=mycolor2,mark=square]
table[x index=0,y index=2]{Plots/2D_UB.dat};

\addplot [color=mycolor3,mark=triangle]
table[x index=0,y index=3]{Plots/2D_UB.dat};

\addplot [color=mycolor4,mark=diamond]
table[x index=0,y index=4]{Plots/2D_UB.dat};

\addplot [color=mycolor5,mark=pentagon]
table[x index=0,y index=5]{Plots/2D_UB.dat};

\addplot [color=mycolor3,mark=triangle*]
table[x index=0,y index=6]{Plots/2D_UB.dat};



\end{axis}
\end{tikzpicture}%

%% file: Plots/3D_UB.tikz
%
%
%
\definecolor{mycolor1}{rgb}{0,0.447,0.741}%
\definecolor{mycolor2}{rgb}{0.85,0.325,0.098}%
\definecolor{mycolor3}{rgb}{0.929,0.694,0.125}%
\definecolor{mycolor4}{rgb}{0.494,0.184,0.556}%
\definecolor{mycolor5}{rgb}{0.466,0.674,0.188}%
\begin{tikzpicture}

\begin{axis}[%
width=\figurewidth,
height=\figureheight,
scale only axis,
separate axis lines,
every outer x axis line/.append style={darkgray!60!black},
every x tick label/.append style={font=\color{darkgray!60!black}},
xmode=log,
xtick={8,16,32,64,128,256,512},
xticklabels={$2^3$,$2^4$,$2^5$,$2^6$,$2^7$,$2^8$,$2^9$},
xlabel={$N^{\frac{1}{3}}$},
xmajorgrids,
every outer y axis line/.append style={darkgray!60!black},
every y tick label/.append style={font=\color{darkgray!60!black}},
ymode=log,
yminorticks=true,
ymajorgrids,
yminorgrids,
legend style={
draw=darkgray!60!black,
fill=white,
legend columns=6,
overlay,
at={(-0.1,1.15)},anchor=south,		
legend cell align=left				
}
]





\addplot [color=mycolor1,mark=o]
table[x index=0,y index=1]{Plots/3D_UB.dat};
\addlegendentry{SPECT};

\addplot [color=mycolor2,mark=square]
table[x index=0,y index=2]{Plots/3D_UB.dat};
\addlegendentry{G2};

\addplot [color=mycolor3,mark=triangle]
table[x index=0,y index=3]{Plots/3D_UB.dat};
\addlegendentry{G4};

\addplot [color=mycolor4,mark=diamond]
table[x index=0,y index=4]{Plots/3D_UB.dat};
\addlegendentry{G6};

\addplot [color=mycolor5,mark=pentagon]
table[x index=0,y index=5]{Plots/3D_UB.dat};
\addlegendentry{G8};

\addplot [color=mycolor3,mark=triangle*]
table[x index=0,y index=6]{Plots/3D_UB.dat};
\addlegendentry{G10};



\end{axis}
\end{tikzpicture}%

%% file: Plots/2D_UB_Grad.tikz
%
%
%
\definecolor{mycolor1}{rgb}{0,0.447,0.741}%
\definecolor{mycolor2}{rgb}{0.85,0.325,0.098}%
\definecolor{mycolor3}{rgb}{0.929,0.694,0.125}%
\definecolor{mycolor4}{rgb}{0.494,0.184,0.556}%
\definecolor{mycolor5}{rgb}{0.466,0.674,0.188}%
\begin{tikzpicture}

\begin{axis}[%
width=\figurewidth,
height=\figureheight,
scale only axis,
separate axis lines,
every outer x axis line/.append style={darkgray!60!black},
every x tick label/.append style={font=\color{darkgray!60!black}},
xmode=log,
xtick={8,16,32,64,128,256,512,1024,2048},
xticklabels={$2^3$,$2^4$,$2^5$,$2^6$,$2^7$,$2^8$,$2^9$,$2^{10}$,$2^{11}$},
xlabel={$N^{\frac{1}{2}}$},
xmajorgrids,
every outer y axis line/.append style={darkgray!60!black},
every y tick label/.append style={font=\color{darkgray!60!black}},
ymode=log,
yminorticks=true,
ylabel={$E_{\infty}$},
ylabel near ticks,
ylabel style={rotate=-90},
ymajorgrids,
yminorgrids,
legend style={
draw=darkgray!60!black,
fill=white,
at={(1.1,0.5)},anchor=west,		
legend cell align=left				
}
]





\addplot [color=mycolor1,mark=o]
table[x index=0,y index=1]{Plots/2D_UB_Grad.dat};

\addplot [color=mycolor2,mark=square]
table[x index=0,y index=2]{Plots/2D_UB_Grad.dat};

\addplot [color=mycolor3,mark=triangle]
table[x index=0,y index=3]{Plots/2D_UB_Grad.dat};

\addplot [color=mycolor4,mark=diamond]
table[x index=0,y index=4]{Plots/2D_UB_Grad.dat};

\addplot [color=mycolor5,mark=pentagon]
table[x index=0,y index=5]{Plots/2D_UB_Grad.dat};

\addplot [color=mycolor3,mark=triangle*]
table[x index=0,y index=6]{Plots/2D_UB_Grad.dat};



\end{axis}
\end{tikzpicture}%

%% file: Plots/3D_UB_Grad.tikz
%
%
%
\definecolor{mycolor1}{rgb}{0,0.447,0.741}%
\definecolor{mycolor2}{rgb}{0.85,0.325,0.098}%
\definecolor{mycolor3}{rgb}{0.929,0.694,0.125}%
\definecolor{mycolor4}{rgb}{0.494,0.184,0.556}%
\definecolor{mycolor5}{rgb}{0.466,0.674,0.188}%
\begin{tikzpicture}

\begin{axis}[%
width=\figurewidth,
height=\figureheight,
scale only axis,
separate axis lines,
every outer x axis line/.append style={darkgray!60!black},
every x tick label/.append style={font=\color{darkgray!60!black}},
xmode=log,
xtick={8,16,32,64,128,256,512},
xticklabels={$2^3$,$2^4$,$2^5$,$2^6$,$2^7$,$2^8$,$2^9$},
xlabel={$N^{\frac{1}{3}}$},
xmajorgrids,
every outer y axis line/.append style={darkgray!60!black},
every y tick label/.append style={font=\color{darkgray!60!black}},
ymode=log,
yminorticks=true,
ymajorgrids,
yminorgrids,
legend style={
draw=darkgray!60!black,
fill=white,
legend columns=6,
overlay,
at={(-0.1,1.15)},anchor=south,		
legend cell align=left				
}
]





\addplot [color=mycolor1,mark=o]
table[x index=0,y index=1]{Plots/3D_UB_Grad.dat};
\addlegendentry{SPECT};

\addplot [color=mycolor2,mark=square]
table[x index=0,y index=2]{Plots/3D_UB_Grad.dat};
\addlegendentry{G2};

\addplot [color=mycolor3,mark=triangle]
table[x index=0,y index=3]{Plots/3D_UB_Grad.dat};
\addlegendentry{G4};

\addplot [color=mycolor4,mark=diamond]
table[x index=0,y index=4]{Plots/3D_UB_Grad.dat};
\addlegendentry{G6};

\addplot [color=mycolor5,mark=pentagon]
table[x index=0,y index=5]{Plots/3D_UB_Grad.dat};
\addlegendentry{G8};

\addplot [color=mycolor3,mark=triangle*]
table[x index=0,y index=6]{Plots/3D_UB_Grad.dat};
\addlegendentry{G10};



\end{axis}
\end{tikzpicture}%

%% file: Plots/Perf1.tikz
%
%
%
\definecolor{mycolor1}{rgb}{0,0.447,0.741}%
\definecolor{mycolor2}{rgb}{0.85,0.325,0.098}%
\definecolor{mycolor3}{rgb}{0.929,0.694,0.125}%
\definecolor{mycolor4}{rgb}{0.494,0.184,0.556}%
\definecolor{mycolor5}{rgb}{0.466,0.674,0.188}%
\begin{tikzpicture}

\begin{axis}[%
width=\figurewidth,
height=\figureheight,
scale only axis,
separate axis lines,
every outer x axis line/.append style={darkgray!60!black},
every x tick label/.append style={font=\color{darkgray!60!black}},
xmode=log,
xtick={1e2,1e3,1e4,1e5,1e6,1e7,1e8},
xticklabels={$10^2$,$10^3$,$10^4$,$10^5$,$10^6$,$10^7$,$10^8$},
xlabel={$N$},
xmajorgrids,
every outer y axis line/.append style={darkgray!60!black},
every y tick label/.append style={font=\color{darkgray!60!black}},
ymode=log,
yminorticks=true,
ylabel={$t_{calc}$ [$\mu s$]},
ylabel near ticks,
ylabel style={rotate=-90},
ymajorgrids,
yminorgrids,
legend style={
draw=darkgray!60!black,
fill=white,
legend pos=north west,
legend cell align=left				
}
]





\addplot [color=mycolor1,mark=o]
table[x index=0,y index=1]{Plots/Perf1.dat};
\addlegendentry{FFTW (1 thread)};

\addplot [color=mycolor2,mark=square]
table[x index=0,y index=2]{Plots/Perf1.dat};
\addlegendentry{FFTW (4 threads)};

\addplot [color=mycolor3,mark=triangle]
table[x index=0,y index=3]{Plots/Perf1.dat};
\addlegendentry{FFTW (8 threads)};

\addplot [color=mycolor4,mark=diamond]
table[x index=0,y index=10]{Plots/Perf1.dat};
\addlegendentry{CUDA};

\addplot [color=black,dashed]
table[row sep=crcr]{
1e4 1e4\\
1e8 1e8\\
};


\end{axis}
\end{tikzpicture}%

%% file: Plots/Bar.tikz
%
%
%
\definecolor{mycolor1}{rgb}{0,0.447,0.741}%
\definecolor{mycolor2}{rgb}{0.85,0.325,0.098}%
\definecolor{mycolor3}{rgb}{0.929,0.694,0.125}%
\definecolor{mycolor4}{rgb}{0.494,0.184,0.556}%
\definecolor{mycolor5}{rgb}{0.466,0.674,0.188}%
\begin{tikzpicture}
    \begin{axis}[
        width  = \figurewidth,
        height = \figureheight,
        major x tick style = transparent,
        ybar=3*\pgflinewidth,
        bar width=14pt,
        ymajorgrids = true,
        ylabel = {Proportion},
        symbolic x coords={ F1,
                            F2,
                            F3,
                            C1,
                            C2,
                            C3},
        xtick = data,
        scaled y ticks = false,
        ymin=0,
        legend cell align=left,
        legend style={
                legend pos=north west,
        }
    ]

                           
        \addplot[style={color=mycolor1,fill=mycolor1,mark=none}]
            coordinates {   (F1,0.484) 
                            (F2,0.464) 
                            (F3,0.484) 
                            (C1,0.639)
                            (C2,0.225)
                            (C3,0.386)};

        \addplot[style={color=mycolor2,fill=mycolor2,mark=none}]
             coordinates {  (F1,0.062) 
                            (F2,0.066) 
                            (F3,0.032) 
                            (C1,0.019)
                            (C2,0.002)
                            (C3,0.03)};

        \addplot[style={color=mycolor3,fill=mycolor3,mark=none}]
             coordinates {  (F1,0.456) 
                            (F2,0.471) 
                            (F3,0.484) 
                            (C1,0.340)
                            (C2,0.772)
                            (C3,0.613)};

        \legend{Forward FFT, Convolution Operations, Backward FFT}
    \end{axis}
    
\end{tikzpicture}%

%% file: SailFFish_Preprint.bbl
\begin{thebibliography}{10}

\bibitem{Gholami_2016}
A.~Gholami, D.~Malhotra, H.~Sundar, and G.~Biros.
\newblock {FFT}, {FMM}, or multigrid? a comparative study of state-of-the-art
  poisson solvers for uniform and nonuniform grids in the unit cube.
\newblock {\em {SIAM} Journal on Scientific Computing}, 38(3):C280--C306, jan
  2016.

\bibitem{Brandt_1982}
A.~Brandt.
\newblock Guide to multigrid development.
\newblock In {\em Multigrid Methods}, pages 220--312, Köln-Porz, 11 1982.

\bibitem{Brandt_1990}
A~Brandt and A.A Lubrecht.
\newblock Multilevel matrix multiplication and fast solution of integral
  equations.
\newblock {\em Journal of Computational Physics}, 90(2):348--370, 1990.

\bibitem{Henson_2003}
Van~Emden Henson.
\newblock {Multigrid methods nonlinear problems: an overview}.
\newblock In Charles~A. Bouman and Robert~L. Stevenson, editors, {\em
  Computational Imaging}, volume 5016, pages 36 -- 48. International Society
  for Optics and Photonics, SPIE, 2003.

\bibitem{Venner_2000}
C.H. Venner and A.A. Lubrecht.
\newblock {\em Multilevel Methods in Lubrication}, volume~37.
\newblock Elsevier Science, 1993.

\bibitem{Adams_1989}
J.C. Adams.
\newblock Mudpack: Multigrid portable fortran software for the efficient
  solution of linear elliptic partial differential equations.
\newblock {\em Appl. Math. \& Comp.}, 34(2):113--146, 1989.

\bibitem{Brandt_2001}
Brandt. A.
\newblock Multiscale scientific computation; review 2001.
\newblock In T.J. Barth, T.~Chan, and R.~Haimes, editors, {\em Multiscale and
  Multiresolution Methods: Theory and Applications}, pages 3--95. Springer
  Berlin, Heidelberg, 1 edition, 2002.

\bibitem{Greengard_1988}
J.~Carrier, L.~Greengard, and V.~Rokhlin.
\newblock A fast adaptive multipole algorithm for particle simulations.
\newblock {\em SIAM Journal on Scientific and Statistical Computing},
  9(4):669--686, 1988.

\bibitem{Greengard_1997}
L.~Greengard and V.~Rokhlin.
\newblock A new version of the fast multipole method for the laplace equation
  in three dimensions.
\newblock {\em Acta Numerica}, 6:229–269, 1997.

\bibitem{Kreyszig_Book}
E.~Kreyszig.
\newblock {\em Advanced Engineering Mathematics}.
\newblock John Wiley \& Sons, 10 edition, 7 2010.

\bibitem{Ying_2004}
Y.~Ying, G.~Biros, and D.~Zorin.
\newblock A kernel-independent adaptive fast multipole algorithm in two and
  three dimensions.
\newblock {\em Journal of Computational Physics}, 196(2):591--626, 2004.

\bibitem{VanGarrel_2017}
A.~van Garrel, C.H. Venner, and H.W. Hoeijmakers.
\newblock Fast multilevel panel method for wind turbine rotor flow simulations.
\newblock In {\em Proc. AIAA SciTech Forum}, Grapevine, Texas, USA, 1 2017.

\bibitem{Saverin-2018-AIAA}
J.~Saverin, D.~Marten, G.~Pechlivanoglou, C.N. Nayeri, and C.O. Paschereit.
\newblock Multilevel simulation of aerodynamic singularity elements.
\newblock In {\em Proc. AIAA SciTech Forum}, Kissimee, Florida, USA, January
  2018.

\bibitem{Cocle_2008}
R.~Cocle, G.~Winckelmans, and G.~Daeninck.
\newblock Combining the vortex-in-cell and parallel fast multipole methods for
  efficient domain decomposition simulations.
\newblock {\em Journal of Computational Physics}, 227(21):9091--9120, 2008.
\newblock Special Issue Celebrating Tony Leonard’s 70th Birthday.

\bibitem{Saverin_PhD}
J.~Saverin.
\newblock {\em Multilevel vortex particle method for aerodynamic simulations}.
\newblock PhD thesis, Technische Universität Berlin, Berlin, Germany, 11 2021.

\bibitem{Cooley_1965}
James~W. Cooley and John~W. Tukey.
\newblock An algorithm for the machine calculation of complex fourier series.
\newblock {\em Mathematics of Computation}, 19:297--301, 1965.

\bibitem{Poplau_2003}
G.~Pöplau and Daniel Potts.
\newblock Fast poisson solvers on nonequispaced grids: Multigrid and fourier
  methods compared.
\newblock {\em Proceedings of SPIE - The International Society for Optical
  Engineering}, 12 2003.

\bibitem{Fuka_2015}
V.~Fuka.
\newblock {PoisFFT} {\textendash} a free parallel fast poisson solver.
\newblock {\em Applied Mathematics and Computation}, 267:356--364, 2015.

\bibitem{Wiegmann_1999}
A.~Wiegmann.
\newblock Fast poisson, fast helmholtz and fast linear elastostatic solvers on
  rectangular parallelepipeds.
\newblock Technical report, Berkeley lab, 6 1999.
\newblock LBNL-43565.

\bibitem{Caprace_2021}
Denis-Gabriel Caprace, Thomas Gillis, and Philippe Chatelain.
\newblock Flups: A fourier-based library of unbounded poisson solvers.
\newblock {\em SIAM Journal on Scientific Computing}, 43(1):C31–C60, Jan
  2021.

\bibitem{Hejlesen_PhD}
M.M. Hejlesen.
\newblock {\em A high order regularisation method for solving the Poisson
  equation and selected applications using vortex method}.
\newblock PhD thesis, DTU Mechanical Engineering, 2016.

\bibitem{Hejlesen_2019}
M.M. Hejlesen, G.~Winckelmans, and J.H. Walther.
\newblock Non-singular green’s functions for the unbounded poisson equation
  in one, two and three dimensions.
\newblock {\em Applied Mathematics Letters}, 89:28--34, 2019.

\bibitem{Orszag_1972}
Steven~A. Orszag.
\newblock Comparison of pseudospectral and spectral approximation.
\newblock {\em Studies in Applied Mathematics}, 51(3):253--259, 1972.

\bibitem{FFTW05}
M.~Frigo and S.~G. Johnson.
\newblock The design and implementation of {FFTW3}.
\newblock {\em Proceedings of the IEEE}, 93(2):216--231, 2005.
\newblock Special issue on ``Program Generation, Optimization, and Platform
  Adaptation''.

\bibitem{FFTW97}
M.~Frigo and S.~G. Johnson.
\newblock The fastest {Fourier} transform in the west.
\newblock Technical Report MIT-LCS-TR-728, Massachusetts Institute of
  Technology, September 1997.

\bibitem{Lackner_1976}
K.~Lackner.
\newblock Computation of ideal mhd equilibria.
\newblock {\em Comp. Phys. Comm.}, 12:33--44, 1976.

\bibitem{James_1977}
R.A James.
\newblock The solution of poisson's equation for isolated source distributions.
\newblock {\em Journal of Computational Physics}, 25(2):71--93, 1977.

\bibitem{Vico_2016}
F.~Vico, Greengard L., and M.~Ferrando.
\newblock Fast convolution with free-space green{\textquotesingle}s functions.
\newblock {\em Journal of Computational Physics}, 323:191--203, oct 2016.

\bibitem{Hockney_1988}
R.W. Hockney and J.W. Eastwood.
\newblock {\em Computer Simulation using Particles}.
\newblock CRC Press, 1988.

\bibitem{Eastwood_1979}
J.W. Eastwood and D.R.K. Brownrigg.
\newblock Remarks on the solution of poisson's equation for isolated systems.
\newblock {\em Journal of Computational Physics}, 32(1):24--38, 1979.

\bibitem{Chatelain_2010}
Chatelain C. and P.~Koumoutsakos.
\newblock A fourier-based elliptic solver for vortical flows with periodic and
  unbounded directions.
\newblock {\em Journal of Computational Physics}, 229(7):2425--2431, 2010.

\bibitem{HejlesenSolver_Github}
M.M. Hejlesen.
\newblock poisson solver.
\newblock \url{https://github.com/mmhej/poissonsolver}, 2019.

\bibitem{OpenMP}
OpenMP Architecture~Review Board.
\newblock Openmp.
\newblock \url{https://www.openmp.org/}, 2021.

\bibitem{FFTW_site}
M.~Frigo and S.~G. Johnson.
\newblock Fftw, 2021.

\bibitem{cufft_site}
NVIDIA Corporation.
\newblock cufft, 2022.

\bibitem{MKL_site}
Intel Corporation.
\newblock Intel mkl.
\newblock
  \url{https://www.intel.com/content/www/us/en/develop/documentation/onemkl-developer-reference-c/top/fourier-transform-functions.html},
  2022.

\bibitem{OpenCLFFT_site}
B.~Natarajan and C.~Robeck.
\newblock clfft.
\newblock \url{https://github.com/clMathLibraries/clFFT}, 2022.

\bibitem{NumpyFFT_site}
NumPy.
\newblock Numpy fft.
\newblock \url{https://numpy.org/doc/stable/reference/routines.fft.html}, 2022.

\end{thebibliography}
